\documentclass[lettersize,journal]{IEEEtran}
\usepackage{amsmath,amsfonts}
\usepackage{algorithmic}
\usepackage{algorithm}
\usepackage{array}
\usepackage[caption=false,font=scriptsize,labelfont=sf,textfont=sf,skip=10pt]{subfig}
\usepackage{textcomp}
\usepackage{stfloats}
\usepackage{url}
\usepackage{verbatim}
\usepackage{graphicx}
\usepackage{cite}
\usepackage{booktabs}
\usepackage{pifont}
\usepackage{float}
\usepackage{multirow,enumitem,hyperref}
\usepackage{soul,color,xcolor,tikz}
\usepackage[normalem]{ulem} 
\usepackage{colortbl}

\usepackage{tikz}
\usetikzlibrary{shapes.geometric}
\newcommand{\sixstar}{%
  \tikz[baseline=-0.5ex]\node[
    star, star points=6, star point ratio=2.2,
    inner sep=0.2ex, minimum size=1.2ex,
    fill=black, draw=none
  ]{};%
}

\begin{document}

\title{Combining Deterministic Enhanced Conditions \\ with Dual-Streaming Encoding \\ for Diffusion-Based Speech Enhancement}

\author{Hao~Shi,~\IEEEmembership{Member,~IEEE},~Xugang~Lu,~\IEEEmembership{Senior Member,~IEEE},~Kazuki~Shimada,~\IEEEmembership{Member,~IEEE},\\~Tatsuya~Kawahara,~\IEEEmembership{Fellow,~IEEE}
\thanks{Hao~Shi,~Kazuki~Shimada,~and~Taysuya~Kawahara are with the Graduate School of Informatics, Kyoto University, Kyoto, Japan (email: shi@sap.ist.i.kyoto-u.ac.jp, Kazuki.Shimada@sony.com, kawahara@i.kyoto-u.ac.jp).}
\thanks{Xugang~Lu is with the National Institute of Information and Communications Technology, Japan (email:xugang.lu@nict.go.jp).}
}

\markboth{Journal of \LaTeX\ Class Files,~Vol.~14, No.~8, August~2021}%
{Shell \MakeLowercase{\textit{et al.}}: A Sample Article Using IEEEtran.cls for IEEE Journals}


\maketitle

\begin{abstract}
Score-based diffusion models for speech enhancement (SE) need to incorporate correct prior knowledge as reliable conditions to generate accurate predictions. 
However, providing reliable conditions using noisy features is challenging. 
One solution is to use features enhanced by deterministic methods as conditions. 
However, the information distortion and loss caused by deterministic methods might affect the diffusion process. 
In this paper, we first investigate the effects of using different deterministic SE models as conditions for diffusion. 
We validate two conditions depending on whether the noisy feature was used as part of the condition: one using only the deterministic feature (deterministic-only), and the other using both deterministic and noisy features (deterministic-noisy). 
Preliminary investigation found that using deterministic enhanced conditions improves DNSMOS and UTMOS on real data, while the choice between using deterministic-only or deterministic-noisy conditions depends on the deterministic models. 
Based on these findings, we propose the deterministic-diffusion unified model for SE to more effectively utilize both conditions. 
Moreover, we found that fine-grained deterministic models have greater potential in objective evaluation metrics, while UNet-based deterministic models provide more stable diffusion performance. 
Therefore, we also propose a deterministic model that combines coarse- and fine-grained processing for the diffusion. 
Experimental results on CHiME4 show that the proposed models effectively leverage deterministic models to achieve better SE evaluation scores on the DNSMOS and UTMOS for real evaluation sets. 
In addition, the proposed deterministic model proves to be more stable than other deterministic models when it is used for diffusion. 
\end{abstract}


\section{Introduction}
\IEEEPARstart{S}{peech} enhancement (SE) is commonly used to recover speech components from noisy speech \cite{5745591,6843279,gao24f_interspeech,dang2024separation,10446645} to improve the speech quality and intelligibility. 
Supervised SE methods \cite{6932438} generally outperform traditional SE methods \cite{5745591,397090} with sufficient training data available. 
These systems can be broadly categorized as either deterministic \cite{10094718,8707065,LI2022108499,10542371,dang23_interspeech} or probabilistic \cite{Pascual2017, 9746901, 10149431,8516711,10448429,shiensemble} approaches. 
Deterministic SE models learn an optimal deterministic mapping from noisy speech to clean speech \cite{10149431}, while probabilistic SE models capture the target distribution in either an implicit or explicit way \cite{Pascual2017, 9746901, 10149431, 8516711}. 
The presence of diverse and uncontrollable sources often makes the noise unpredictable, resulting in random signals. 
However, deterministic SE models only learn a deterministic mapping function, leading to performance degradation in unseen noise scenarios.

Diffusion model \cite{kamo2023target,universal_2022,10149431,9746901,shiensemble,10248171} is a probabilistic method learning a stochastic differential equation (SDE). 
It contains a forward diffusion process and a reverse diffusion process by following the SDE. 
The forward diffusion process gradually transforms the clean speech feature into noise, during which the neural network learns to reverse the incremental process of noise addition. 
The reverse process recovers the speech feature from noise with numerous reverse iterations. 
Different from the deterministic models \cite{10094718,8707065,LI2022108499,10542371}, SDE has the characteristics of the partially stochastic as: random noise is injected into each reverse SDE step.  
Score-based diffusion model for SE \cite{9746901, 10149431, 10180108} takes the clean speech feature as the initial state and transforms it to the noisy speech feature with Gaussian white noise. 
Noisy speech feature is used as condition \cite{9746901, 10149431, 10180108} and drift term \cite{10149431, 10180108} to enable the model to capture noise information implicitly. 
However, noisy features pose a challenge in providing accurate prior knowledge as reliable conditions for diffusion. 
Schr\"odinger Bridge-based diffusion model learns the diffusion process directly between the noisy input and the clean distribution, rather than relying on Gaussian reference distributions, and it also achieves high performance in SE \cite{jukic24_interspeech}.

Combining deterministic and diffusion models \cite{10096985, 10180108, 10448429, 10832147} makes a promising solution for alleviating this challenge. 
Both noisy and deterministically enhanced (deterministic-noisy) features \cite{10832147} or only deterministically enhanced (deterministic-only) features are used as conditions for diffusion \cite{10180108}. 
When using a diffusion model to process deterministic enhanced features, the posterior samples are more likely to lie in high-density regions of the posterior probability for clean speech features given noisy speech features\cite{10180108}. 
The effectiveness of the combination has been demonstrated \cite{10180108, 10096985, 10832147}, but no previous work has investigated on how different deterministic models affect the SE performance based on this pipeline.

In this paper, we investigate the effects of using different deterministic SE models \cite{hu20g_interspeech, 8707065, 9054266, 10214650} as conditions for diffusion-based SE. 
We experiment with score-based diffusion models that use either deterministic-only enhanced features (Deter-Diffusion) or deterministic-noisy features (StoRM \cite{10180108} and GP-Unified \cite{10448429}) as conditions. 
Deter-Diffusion and StoRM train the diffusion model only based on the score-matching loss, without incorporating any deterministic loss. 
Some studies have shown that deterministic information and diffusion-based information are complementary \cite{universal_2022,10448429}. 
Incorporating deterministic information—either from the encoder \cite{10448429} or the decoder \cite{universal_2022}—into diffusion models enhances performance. 
Rather than relying on complex fusion operations in the decoder \cite{universal_2022}, simply introducing the deterministic loss after the encoder during training also yields significant performance improvements \cite{10448429}. 
GP-Unified \cite{10448429} adopts the same encoder-decoder architecture as StoRM \cite{10180108}. 
In addition to the score error, GP-Unified \cite{10448429} introduces deterministic information into the diffusion model using an additional decoder. 
Although GP-Unified \cite{10448429} outperforms StoRM \cite{10180108}, performing both deterministic estimation and diffusion within a shared bottleneck processing module may lead to performance degradation under noise-mismatch conditions \cite{10248171}, as both the deterministic and score-estimation information exhibit mismatched data distributions between simulated and real data.

\begin{figure}
    \centering
    \includegraphics[width=0.9\linewidth]{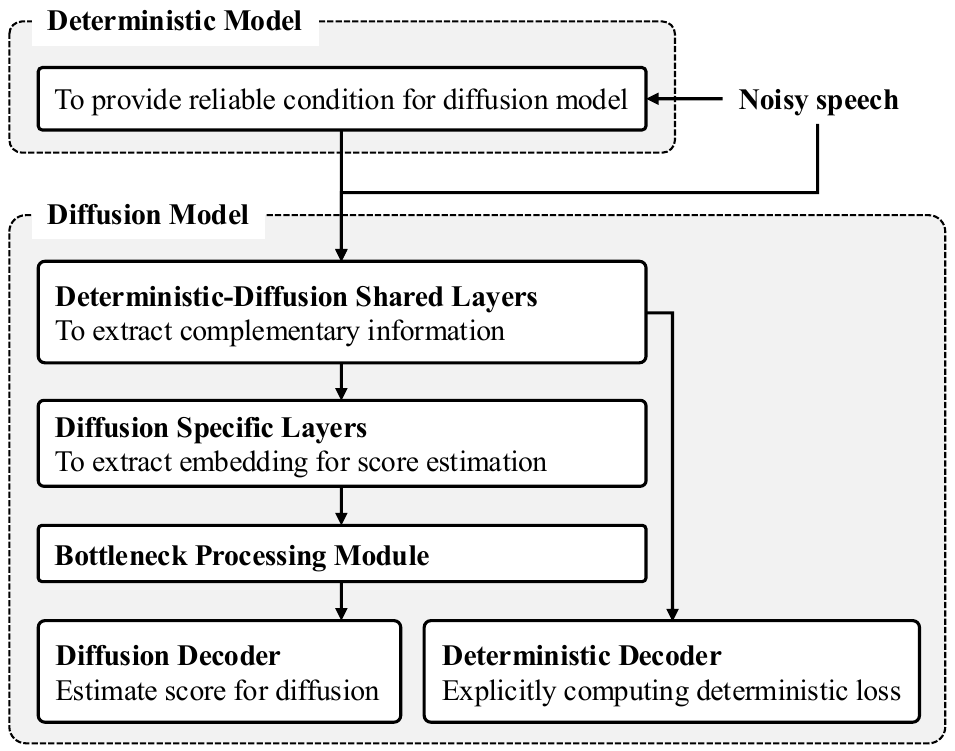}
    \caption{High-level conceptual flowchart of the D2U-SE framework. 
    Each stage corresponds to a core modeling motivation.}
    \label{fig:abstract-flowchart}
\end{figure}

To more effectively utilize deterministic information as conditioning during diffusion, we propose a Deterministic-Diffusion Unified model for SE (D2U-SE), which explicitly incorporates the deterministic loss during the training of the diffusion model. 
The key difference from GP-Unified \cite{10448429} is that D2U-SE incorporates deterministic information from only a subset of the encoder layers rather than all encoder layers and the shared bottleneck module. 
Recent studies suggest that feature downsampling may hinder further improvement of deterministic SE performance \cite{8707065,9054266, 10214650}. 
Thus, we adopt different downsampling strategies across the encoder layers to better incorporate deterministic information and extract diffusion embeddings. 
In the shallow encoder layers, we perform downsampling only along the frequency axis to better extract deterministic information. 
After getting the deterministic-diffusion embedding, the diffusion-specific layers perform downsampling along the time axis or the time-frequency axis to extract embeddings better suited for score estimation. 
Fig.~\ref{fig:abstract-flowchart} shows the high-level flowchart of the D2U-SE. 
The choice between deterministic-only and deterministic-noisy conditions varies across deterministic models. 
We further enhance the D2U-SE with Dual-streaming encoding (D3U-SE) to more effectively utilize both deterministic-only and deterministic-noisy conditions.

Although fine-grained processing deterministic models, such as TF-GridNet \cite{10214650}, demonstrate strong enhancement performance on simulated data, diffusion models with fine-grained enhanced deterministic conditions show less stable performance on real speech compared to the coarse-grained enhancement deterministic models with a UNet structure, such as DCCRN \cite{hu20g_interspeech}. 
Thus, we propose a granularity progressive deterministic model, called {CO}arse {F}isrt then {F}ine {E}nhanc{E}ment (COFFEE), that combines fine-grained processing with the UNet structure. 
It enhances the feature starting from the bottleneck embedding, progressing to the reconstructed feature, and finally to the time-frequency bin using fine-grained processing.

In the following sections, we will introduce the preliminaries in Section~\ref{preliminaries}. 
We will explain the proposed method in Section~\ref{proposed_method}. 
The experimental settings and results are presented in Section~\ref{experimental}. 
The conclusion will be given in Section~\ref{conclusions}.

\section{Preliminaries}
\label{preliminaries}
\subsection{Deterministic Methods for Speech Enhancement}
The UNet structure has been successfully applied in SE, either in frequency-domain or waveform-domain methods \cite{Dfossez2020,hu20g_interspeech}. 
UNet-based SE models consist of an encoder, a decoder, and a bottleneck processing module. 
With its downsampling layers, the encoder plays a crucial role in feature extraction and dimensionality reduction. 
The decoder upsamples the feature maps and reconstructs the enhanced feature from the encoded feature representations to the original dimensions. 
The bottleneck processing module captures the abstract and compressed representation from the encoder.

Recent studies suggest that feature downsampling may hinder further improvement of SE performance \cite{8707065,9054266, 10214650}. 
The TasNet-series \cite{8707065,9054266} have attracted much attention due to their excellent separation ability. 
They use a separator to process the pseudo-spectrogram extracted by the Conv1d layer. 
In neither the Conv-TasNet with the temporal convolutional network (TCN) nor the dual-path RNN (DPRNN) are the features rarely downsampled in the separator blocks. 
Some related works also show that extracting features with smaller strides performs better. 
With time-frequency (T-F) bin-level enhancement, TF-GridNet \cite{10214650} shows more powerful performance and achieves state-of-the-art performance in many commonly used datasets. 
It does not do any downsampling processing. 
It first extracts the complex spectrogram; then, each T-F bin is enhanced twice along the time-domain and frequency-domain axes with several TF-GridNet blocks. 
With the small granularity processing, the processing speed is extremely slow.

\label{sec:tfgridnet}
TF-GridNet performs complex spectral mapping. 
The stacked real and imaginary (RI) parts of the complex noisy spectrogram $\mathbf{Y} = \mathbf{Y}_r + j\mathbf{Y}_i \in\mathbb{C}^{T\times F}$ are used as input features, where $T$ and $F$ represent the spectrogram's time and frequency. 
$j$ is the imaginary unit as $j^2=-1$. 
Each T-F bin first extracts a $D$-dimensional embedding with a Conv2D (kernel size $=3 \times 3$ ) layer. 
The extracted feature is $R_a \in \mathbb{R}^{D \times T \times F}$. 
Then, several TF-GridNet blocks enhance the noisy embeddings. 
An intra-frame full-band module, a sub-band temporal module, and a cross-frame self-attention module form the TF-GridNet block. 
The complex spectrogram embedding is first zero-padded to the shape $R_b \in D \times T \times F'$ to ensure that adjacent information can be obtained when processing each T-F bin, where $ F' = \left\lfloor \frac{F - I}{J} \right\rfloor \times J + I $. 
$I$ and $J$ represent the kernel size and stride, respectively. 
In the intra-frame full-band module, $R_b$ is first unfolded as $R_p$: 
\begin{equation}
\begin{aligned}
R_p &= \left[\text{Unfold}\left(\text{LN}(R_b)[:, t, :]\right), \right. \left. \text{for } t = 1, \ldots, T \right]
\end{aligned}
\end{equation}
$\text{LN}$ represents the layer normalization. 
A single bi-directional long short-term memory network (referred to as \textbf{Intra-LSTM}), with $H$ hidden nodes, is used to process the unfolded feature $R_p$ along the time axis of the spectrogram: 
\begin{equation}
\begin{aligned}
R_t &= \left[\text{BLSTM}\left(\text{LN}(R_p)[:, t, :]\right), \right. \left. \text{for } t = 1, \ldots, T \right]
\end{aligned}
\end{equation}
Then, a DeConv1D layer with kernel size $I$, stride $J$, input channel $2H$ (Bi-directional LSTM is used here), and output channel D (and without subsequent normalization and non-linearity) is applied to full-band processed embedding $R_t$: 
\begin{equation}
\begin{aligned}
R_d &= \left[\text{DeConv1D}(R_t[:, t, :]), \right. \left. \text{for } t = 1, \ldots, T \right]
\end{aligned}
\end{equation}
$R_d$ is added to the input tensor after removing zero paddings: 
\begin{equation}
U_b = R_d[:, :, :F] + R_b[:, :, :F] \in \mathbb{R}^{D \times T \times F}
\end{equation}
The sub-band temporal module has the same structure as the intra-frame full-band module, except that the \textbf{Intra-LSTM} is replaced with \textbf{Inter-LSTM}. 
It processes the input tensor along the frequency axis of the spectrogram, which obtains the sub-band enhanced embedding $Z_b \in \mathbb{R}^{D \times T \times F}$.

In the cross-frame self-attention module, the frame embeddings are first computed. 
Then, the full-utterance self-attention is used to model long-range context information. 
The self-attention module has $L$ heads. 
The point-wise Conv2D, PReLU, and layer normalization along the channel and frequency dimensions (cfLN) are used in each head to obtain query, key, and value. 
Then, the query, key, and value are reshaped as $Q_l \in \mathbb{R}^{T \times (F \times E)}$, $K_l \in \mathbb{R}^{T \times (F \times E)}$ and $V_l \in \mathbb{R}^{T \times (F \times D / L)}$. 
The point-wise Conv2D channels for computing the query and key are $E$, while the value is $D / L$. 
All three point-wise Conv2D layers have $D$ input channels. 
The attention outputs of all heads along the second dimension are concatenated and reshaped back to $D \times T \times F$. 
Another combination of the point-wise Conv2D with $D$ input and $D$ output channels, followed by a PReLU and a cfLN, is applied to aggregate cross-head information. 
Finally, the residual connection of the $Z_b$ is added to the output of the attention module.

After several TF-GridNet blocks processing, a 2D deconvolution (DeConv2D) with a $3 \times 3$ kernel followed by linear units is used to obtain the predicted RI components of the enhanced speech $\tilde{\mathbf{X}}_0 = \tilde{\mathbf{X}}_{0r} + j\tilde{\mathbf{X}}_{0i} \in\mathbb{C}^{T\times F}$. 
The training loss is based on the Mean Squared Error (MSE) between the clean and enhanced spectrogram of their stacked RI parts: 
\begin{equation}
	\mathcal{L}_{Deter} =  \left\| \tilde{\mathbf{X}}_0 - \mathbf{X}_0 \right\|^2_2 
	\label{deterloss}
\end{equation}
where ${\mathbf{X}}_0 = {\mathbf{X}}_{0r} + j{\mathbf{X}}_{0i} \in\mathbb{C}^{T\times F}$ represents the clean spectrogram.

\subsection{Score-based Diffusion Model for Speech Enhancement}
Diffusion model \cite{NEURIPS2019_3001ef25,NEURIPS2021_b578f2a5,NEURIPS2021_c11abfd2,ANDERSON1982313} belongs to the probabilistic method \cite{NEURIPS2020_4c5bcfec}. 
The diffusion model has two processes: the forward process and the reverse process. 
The forward process follows the stochastic differential equation (SDE). 
A clean speech feature $\mathbf{x}_0$ is gradually transformed to a noisy speech feature $\mathbf{y}$ and white Gaussian noise \cite{10149431,kamo2023target} with linear SDE: 
\begin{equation}
    \mathrm{d}\mathbf{\mathbf{X}}_t = \mathbf{f}(\mathbf{X}_t, \mathbf{Y}) \mathrm{d}t + g(t) \mathrm{d}\mathrm{\mathbf{w}},
    \label{eq:forward_sde}
\end{equation}
where $\mathbf{X}_t$ is the current state, and $t$ is a continuous state index $[0, T]$. 
$\mathrm{\mathbf{w}}$ represents a standard Wiener process, which is also called Brownian motion. 
$\mathbf{f}(\mathbf{X}_t, \mathbf{Y})$ is the \textit{drift coefficient}, while $g(t)$ is the \textit{diffusion coefficient}. 
The forward SDE has a corresponding reverse SDE, which is used to achieve SE \cite{song2021scorebased,NEURIPS2021_c11abfd2}. 
The score $\nabla_{\mathbf{X}_{t}}\mathrm{log} p_{t}(\mathbf{X}_t|\mathbf{X}_0, \mathbf{Y})$ on a state of the SDE needs to be estimated for reverse SDE \cite{song2021scorebased,ANDERSON1982313}. 
A neural network $s_\theta$ with parameter set $\theta$, called the scoring model, is used to approximate the score. 
The noisy speech feature $\mathbf{Y}$ is used as the condition and makes the conditional probability as $p_t(\mathbf{X}_{t}|\mathbf{X}_0, \mathbf{Y})$. 
Thus, the inputs of the score models are defined as $s_{\theta}(\mathbf{X}_{t},\mathbf{Y},t)$. 
The reverse SDE is defined as follows: 
\begin{equation}
\mathrm{d}\mathbf{X}_{t} = \left [ -\mathbf{f}(\mathbf{X}_t, \mathbf{Y}) + g(t)^{2}s_{\theta} \right ] \mathrm{d}t + g(t)\mathrm{d}\bar{\mathrm{\mathbf{w}}}, 
\label{r_sde_eq}
\end{equation}
The SE is achieved by solving the reverse SDE starting from $t=T$ back to $t=0$ iteratively. 
At $t$, the state $\mathbf{x}_t$ for the reverse process is sampled as follows \cite{10149431}: 
\begin{equation}
	\mathbf{X}_t \sim \mathcal{N}_{c} (\mathbf{X}_t; \mathbf{Y}, \sigma (t)^2\mathbf{I} ),
    \label{eq:init}
\end{equation}
where $\mathcal{N}_{c}$ represents the circularly-symmetric complex normal distribution. 
$\mathbf{I}$ represents the identity matrix. 
$\sigma (t)^2$ is the variance of the white Gaussian noise included in $\mathbf{X}_{t}$ according to the forward SDE: 
\begin{equation}
    \sigma (t)^2 = \frac{c(k^{2t} - e^{-2\gamma t})}{2(\gamma + \log k)},
    \label{sigma_equation}
\end{equation}
where $k$, $\gamma$, and $c$ are positive scalar constants.

The overall training objectives for denoising are derived as follows: 
\begin{equation}
	\underset{\theta }{\mathrm{arg}~\mathrm{min}\,} \mathbb{E}\left [ \left\|  \mathrm{s}_\theta +  \mathrm{z}/{\sigma (t)}  \right\|^2_2 \right ],  
	\label{score_objects}
\end{equation}
where $\mathrm{z} \sim \mathcal{N_C}(\mathrm{z}; 0, \mathbf{I})$ is a sampled white Gaussian noise with an identity matrix $\mathbf{I}$. 
The loss function is defined as denoising score matching: 
\begin{equation}
    \mathcal{L}_{\mathrm{Score}} = \lambda(t) \left\| s_\theta(\mathbf{X}_t, \mathbf{Y}, t) - \nabla_{\mathbf{X}_t} \log p_{t}(\mathbf{X}_t | \mathbf{X}_0, \mathbf{Y}) \right\|_2^2,
	\label{score_loss}
\end{equation}
where $\lambda(t)$ is the weighting function  that depends on $t$.

The interval $[0, T]$ is partitioned into $N$ steps of width $\Delta t = T/N$ to find the solution of the reverse SDE numerically. 
The discrete reverse process over $\{X_T, X_{T-\Delta t}, ..., X_0 \}$ is utilized.
The Predictor-Corrector (PC) sampler \cite{song2021scorebased} is used to solve the reverse SDE. 
At each reverse diffusion step, the current state $\mathbf{X}_{t}$ is determined from the previous step $\mathbf{X}_{t+1}$ by applying both predictor and corrector. 
Each PC sampling needs to call the score model $s_{\theta}$ at least once. 
At each reverse diffusion step, White Gaussian noise is introduced to both predictor and corrector according to Eq.~(\ref{r_sde_eq}). 
As a result, each reverse SDE process follows a distinguished Brownian motion trajectory.

\subsection{The Relationship Between Score-based Diffusion Model and Deterministic Model for Speech Enhancement}
Based on the formulation in Eq.~(\ref{score_loss}), denoising score matching can be expressed as training a denoiser model $D_\theta$ to match the clean signal estimate by \cite{10887784,NEURIPS2022_a98846e9}: 
\begin{equation}
\begin{aligned}
    \mathcal{L}_{\text{Score-denoiser}} 
    = \lambda(t) \left\| D_\theta(\mathbf{X}_t, \mathbf{Y}, t) - \mu_t(\mathbf{X}_0, \mathbf{Y}) \right\|_2^2 \\
    = \lambda(t) \left\| D_\theta(\mathbf{X}_t, \mathbf{Y}, t) - (1 - e^{-\gamma t}) \mathbf{Y} - e^{-\gamma t} \mathbf{X}_0 \right\|_2^2.
\end{aligned}
\end{equation}
When $t = 0$, 
\begin{equation}
\mathcal{L}_{\text{Score-denoiser}} = \lambda(t) \left\| D_\theta(\mathbf{X}_t, \mathbf{Y}, t) - \mathbf{X}_0 \right\|_2^2. 
\label{lossscoredenoiser}
\end{equation}
Similarly, the training loss of the deterministic model can also be expressed in a similar format based on Eq.~(\ref{deterloss}): 
\begin{equation}
    \mathcal{L}_{\text{Deter-denoiser}} = \left\| D_\theta(\mathbf{Y}) - \mathbf{X}_0 \right\|_2^2.
\label{lossdeterdenoiser}
\end{equation}
Compared with Eq.~(\ref{lossscoredenoiser}) and Eq.~(\ref{lossdeterdenoiser}), both the deterministic and score-based diffusion models aim to recover the clean feature from its noisy feature. 
The key differences lie in the presence of the time-dependent scaling factor and in whether the noisy feature $\mathbf{Y}$ is explicitly incorporated during the denoising process. 
Besides, the choice of the weighting function $\lambda(t)$ can have a substantial impact in practice and may lead to slight performance differences.

\section{Proposed Method}
\label{proposed_method}
\begin{figure*}
    \centering
    \includegraphics[width=0.9\textwidth]{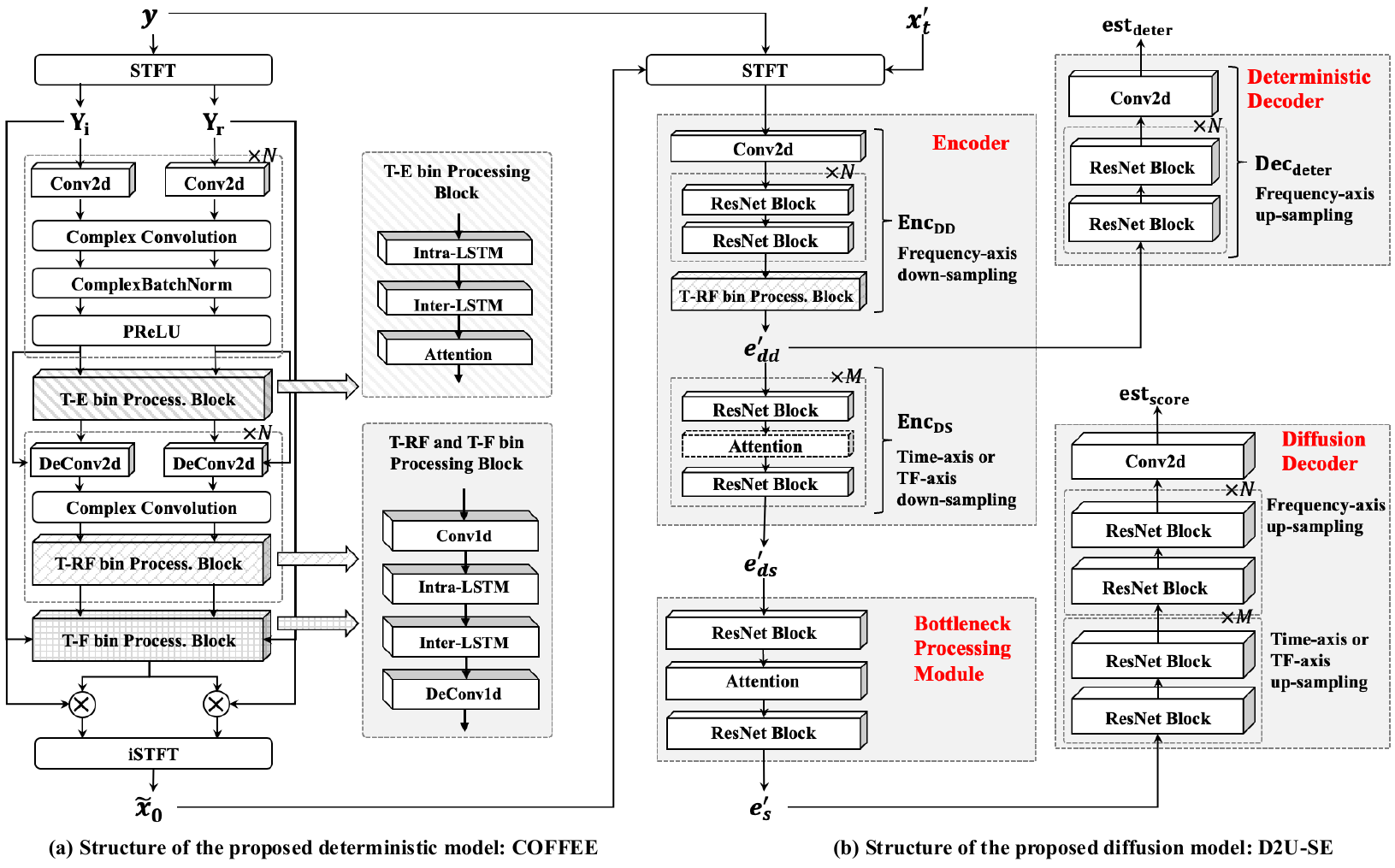}
    \caption{Flowchart of the proposed method. 
    It includes the proposed COFFEE deterministic model and the proposed D2U-SE. 
    Within COFFEE, we introduce Time-Embedding (T-E) bin-level, Time-Reconstructed Feature (T-RF) bin-level, and Time-Frequency (T-F) bin-level processing. 
    Within D2U-SE, the skip connections between the encoder and decoder are omitted in the figure for clarity.
    Lowercase letters denote the time-domain representations of their corresponding uppercase symbols.
    }
    \label{fig:pipeline}
\end{figure*}

We propose combining the deterministic and diffusion models for SE. 
It contains a deterministic model and a diffusion model as shown in Fig.~\ref{fig:pipeline}. 
In the deterministic model, we propose a granularity progressive deterministic model COFFEE by leveraging the advantages of both high-speed down-sampling processing and high-performance fine-granularity processing as shown in Fig.~\ref{fig:pipeline}. 
The bottleneck embedding, reconstructed features in the decoder, and the T-F bins are enhanced with fine-granularity processing gradually.

To better integrate the deterministic and diffusion models, we propose the D2U-SE. 
We design the encoder of D2U-SE based on the characteristics of deterministic and diffusion processes. 
Within the encoder, the features are processed through several deterministic-diffusion layers and diffusion-specific layers. 
Deterministic information is explicitly introduced into the shared encoder layers by computing the deterministic loss. 
The deterministic-diffusion shared layers perform downsampling only along the frequency axis to retain sufficient information for learning deterministic information. 
The diffusion-specific layers focus on extracting embeddings only for diffusion. 
Besides, the deterministic models affect the choice between using deterministic-only and deterministic-noisy conditions. 
Thus, we further improve the D2U-SE with dual-stream encoding (D3U-SE) to simultaneously utilize both deterministic-only and deterministic-noisy conditions.

\subsection{Deterministic Method: COFFEE}
We use a complex-valued network, which often requires decomposing a signal into its real and imaginary components, potentially avoiding the loss of important phase information and enhancing performance in certain tasks \cite{bassey2021surveycomplexvaluedneuralnetworks}. 
It contains an encoder, a decoder, and the bottleneck processing layers. 
The encoder reduces the resolution of the input feature and extracts high-level representations. 
The encoder layer processes the input complex feature $\mathbf{Y} = \mathbf{Y}_r + j\mathbf{Y}_i \in\mathbb{C}^{T\times F}$ with complex Conv2d. 
The complex-valued filter $W = W_r + jW_i$ is used for convolution operation: 
\begin{equation}
\mathbf{Y}_{o} = (\mathbf{Y}_r * W_r^e - \mathbf{Y}_i * W_i^e) + j(\mathbf{Y}_r * W_i^e + \mathbf{Y}_i * W_r^e), 
\end{equation}
The real-valued matrices $W_r^e$ and $W_i^e$ represent the real and imaginary parts of the complex convolution kernel, respectively. 
$W_r^e$ and $W_i^e$ are implemented with the Conv2d layer with different parameters. 
$\mathbf{Y}_{o}$ is the filtered feature. 
Batch normalization and PReLU follow the Conv2d layers. 
\begin{equation}
\begin{bmatrix}
\hat{\mathbf{Y}_{o}}_r \\
\hat{\mathbf{Y}_{o}}_i
\end{bmatrix}
=
\operatorname{PReLU}(W^{a}
\cdot
V^{-\frac{1}{2}}
\cdot
\begin{bmatrix}
\mathbf{Y}_{or} - \mathbb{E}[\mathbf{Y}_{or}] \\
\mathbf{Y}_{oi} - \mathbb{E}[\mathbf{Y}_{oi}]
\end{bmatrix}
+
\begin{bmatrix}
\mathbf{b}_r \\
\mathbf{b}_i
\end{bmatrix})
\end{equation}
where $\mathbf{Y}_{or}$ and $\mathbf{Y}_{oi}$ are the real and imaginary parts of $\mathbf{Y}_{o}$, respectively. 
The mean $\mathbb{E}[\cdot]$ is computed over the batch dimension. 
The matrix $V^{-1/2} \in \mathbb{R}^{2 \times 2}$ denotes the inverse square root of the $2 \times 2$ covariance matrix constructed from the real and imaginary components. 
The learnable affine transformation is represented by $W^a$, and the bias vector by $\mathbf{b}_r$ and $\mathbf{b}_i$. 
Finally, $\operatorname{PReLU}(\cdot)$ denotes the element-wise parametric ReLU activation function applied separately to each component.

The strategy for the bottleneck Time-Embedding (T-E) bins processing is that the real and imaginary parts are modeled jointly: 
\begin{equation}
B_{out} = B * W^b, 
\end{equation}
where the $W^b$ is implemented with the T-E processing block. 
$B$ is the embedding extracted by the encoder. 
Fig.~\ref{fig:pipeline} shows the structure of the T-E processing Block. 
It contains an Intra-LSTM, an Inter-LSTM, and an attention mechanism to enhance full-band and sub-band modeling capabilities. 
Two additional linear layers with complex convolution are included for separate modeling. 
The Intra-LSTM, Inter-LSTM, and Attention are similar to the process of processing features described in Section~\ref{sec:tfgridnet}. 
The T-E bin level feature is processed with a large dimension reduction compared with the original T-F spectrogram.

The decoder reconstructs the original resolution feature from the bottleneck embedding. 
In each decoder layer, the features are processed as follows: 
\begin{equation}
    D_{out} = (D_r * W_r^d - D_i * W_i^d) + j(D_r * W_i^d + D_i * W_r^d), 
\end{equation}
$W_r^d$ and $W_i^d$ are implemented with the ConvTranspose2d layer. 
A Time-Reconstructed Feature (T-RF) processing Block is used to process the reconstructed feature after the separated modeling of the real and imaginary parts of T-RF. 
The structure of the T-RF processing block is shown in Fig.~\ref{fig:pipeline}. 
It contains a Conv2d layer, an Intra-LSTM, an Inter-LSTM, and a DeConv2d layer. 
The Conv1d layer aims to extract the embedding for each T-RF bin. 
The time and frequency of the reconstructed feature remain unchanged. 
The T-RF bin level feature still has a dimension reduction compared with the original T-F spectrogram. 
The DeConv1d layer is used to recover the embedding dimension of each T-RF bin. 
The time and frequency dimensions of the reconstructed feature remain unchanged.

Finally, after the feature is upsampled to the same resolution as the original noisy feature, a T-F processing Block is used to make the final refinement. 
The T-F processing block has the same structure as the T-RF processing block.

Signal approximation (SA) is used to train the complex ratio mask (CRM) of COFFEE. 
CRM is defined as follows:
\begin{equation}
CRM = \frac{\mathbf{Y}_r\mathbf{X}_{0r} + \mathbf{Y}_i\mathbf{X}_{0i}}{\mathbf{Y}_r^2 + \mathbf{Y}_i^2} + j \frac{\mathbf{Y}_r\mathbf{X}_{0i} - \mathbf{Y}_i\mathbf{X}_{0r}}{\mathbf{Y}_r^2 + \mathbf{Y}_i^2}
\end{equation}
where $\mathbf{Y}_r$ and $\mathbf{Y}_i$ represent the real and imaginary parts of the noisy complex spectrogram $\mathbf{Y}$, respectively. 
The $\mathbf{X}_{0r}$ and $\mathbf{X}_{0i}$ represent the real and imaginary parts of the clean complex spectrogram $\mathbf{X}_{0}$, respectively. 
According to the polar coordinates of Cartesian coordinate representation $\tilde{M} = \tilde{M}_r + j\tilde{M}_i$: 
\begin{equation}
\left\{
\begin{aligned}
\tilde{M}_{\text{mag}} &= \sqrt{\tilde{M}_r^2 + \tilde{M}_i^2}, \\
\tilde{M}_{\text{phase}} &= \arctan 2(\tilde{M}_i, \tilde{M}_r).
\end{aligned}
\right.
\end{equation}
The estimated enhanced spectrogram can be calculated as follows:
\begin{equation}
 \tilde{\mathbf{X}}_0 = \mathbf{Y}_{mag} \cdot \tilde{M}_{mag} \cdot e^{j*(\mathbf{Y}_{phase} + \tilde{M}_{phase})}
\end{equation}
where $\mathbf{Y}_{mag}$ and $\mathbf{Y}_{phase}$ represent the magnitude and phase information of the noisy speech. 
The training loss is based on the MSE between the clean and enhanced spectrogram, which is the same as in Eq.~(\ref{deterloss})

\subsection{Diffusion Model: D2U-SE}
To alleviate the effect of the environmental noise, a deterministic model, such as COFFEE, is first used to process the noisy speech feature: 
\begin{equation}
   \mathbf{X'} = \underset{\text{COFFEE}}{\text{Deterministic}}(\mathbf{Y})
\end{equation}
where $\mathbf{X'}$ is the enhanced speech with deterministic model. 
For the diffusion model, we follow the reverse SDE, which is similar to Eq.~(\ref{r_sde_eq}): 
\begin{equation}
\mathrm{d}\mathbf{X}_{t}' = \left [ -\mathbf{f}(\mathbf{X}_{t}', \mathbf{X'}, \mathbf{Y}) + g(t)^{2}s_{\theta} \right ] \mathrm{d}t + g(t)\mathrm{d}\bar{\mathrm{\mathbf{w}}}. 
\label{r_deter_sde_eq}
\end{equation}
The deterministic enhanced and noisy features are used as the condition. 
Thus, the $\nabla_{\mathbf{X}_{t}'}\mathrm{log} p_{t}(\mathbf{X}_{t}'|\mathbf{X'}, \mathbf{Y})$ on a state of the SDE is needed to be estimated for reverse SDE \cite{song2021scorebased,ANDERSON1982313}. 
The previous works found a solution of the reverse SDE numerically within the interval $[0, T]$ (partitioned into $N$ steps of width $\Delta t = T/N$). 
Different from the previous works \cite{10180108, 10096985}, the proposed method starts the diffusion from a middle state $\mathbf{X}_{k}' (k \in [0, T])$ instead of from the very initial state $\mathbf{X}_{T}'$ since the input speech is already enhanced and the diffusion model is used for refining. 
The $\mathbf{X}_{k}'$ can be obtained according to the forward SDE: 
\begin{equation}
    \mathrm{d}\mathbf{\mathbf{X}}'_{k} = \mathbf{f}(\mathbf{X}'_{k}, \mathbf{X'}, \mathbf{Y}) \mathrm{d}k + g(k) \mathrm{d}\mathrm{\mathbf{w}},
\end{equation}
The following diffusion steps are the same as the original diffusion. 
We utilize the property of Brownian motion to augment the enhanced signals following $M$ different reverse SDEs: 
\begin{equation}
    \mathbf{X}'_{0m} = \text{Diffusion}(\mathbf{X}'_{Km}, \mathbf{X'}, \mathbf{Y})
\end{equation}
where $\mathbf{X'_{0}}_m$ is the $m$-th diffusion enhanced speech feature. 
The final enhanced signal is obtained by ensembling $M$ augmented enhanced speech signals to utilize the properties of distinct Brownian motion trajectories \cite{kamo2023target, shiensemble}: 
\begin{equation}
    \overline{\mathbf{X}'_{0m}} = {1}/{M} \textstyle \sum_{m=1}^{M} \mathbf{X}'_{0m}.
    \label{ensemble}
\end{equation}

\begin{figure}
    \centering
    \includegraphics[width=0.45\textwidth]{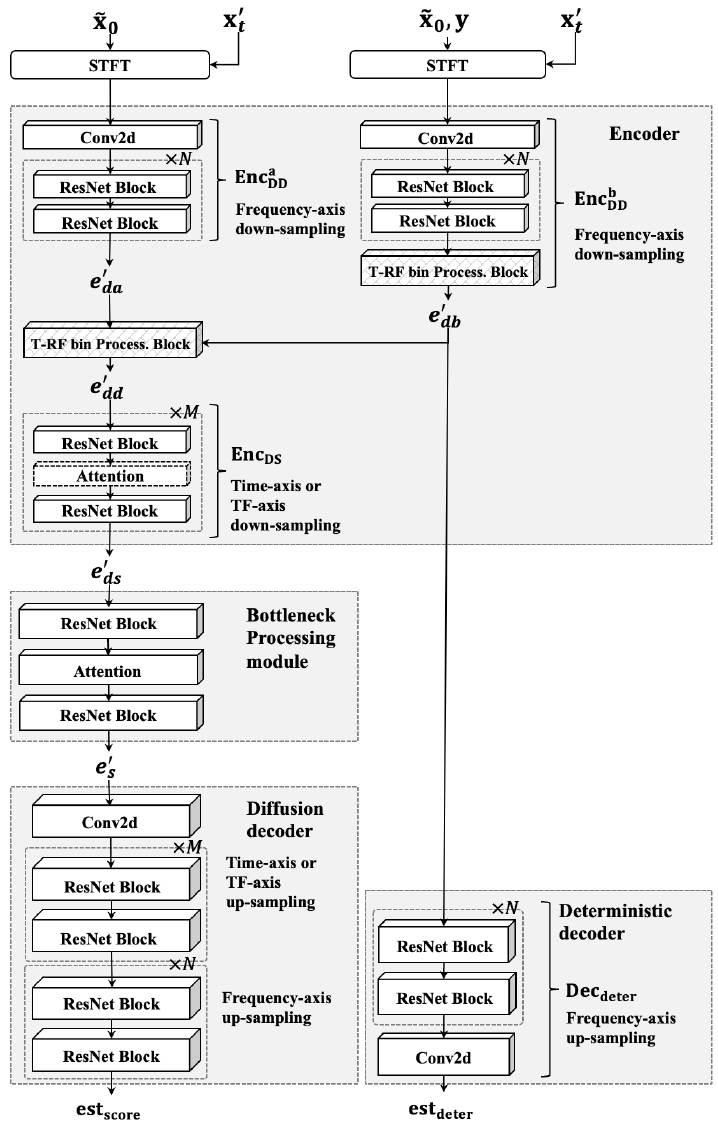}
    \caption{Flowchart of the proposed D3U-SE. 
    The skip connections between the encoder and decoder are omitted in the figure for clarity.
    Lowercase letters denote the time-domain representations of their corresponding uppercase symbols.}
    \label{fig:dual-streaming}
\end{figure}

We propose the D2U-SE to better utilize deterministic information. 
It is based on a dual-decoder structure. 
The model contains an encoder, a bottleneck processing module, a diffusion decoder, and a deterministic decoder. 
Within the encoder, the feature is processed by several deterministic-diffusion shared layers and diffusion-specific layers. 
The conditions and diffused features are first processed by the encoder: 
\begin{equation}
\begin{cases}
   \mathbf{e'_{dd}} = \text{Enc}_\text{DD}(\mathbf{X'}_t, \mathbf{X'}, \mathbf{Y}, t) \\ 
   \mathbf{e'_{ds}} = \text{Enc}_\text{DS}(\mathbf{e'_{dd}}, t),
\end{cases}
\label{rdm_encoder}
\end{equation}
where $\text{Enc}_\text{DD}$ and $\text{Enc}_\text{DS}$ represent the deterministic-diffusion shared layers and diffusion-specific layers, respectively. 
$\mathbf{e'_{dd}}$ and $\mathbf{e'_{ds}}$ denote the deterministic-diffusion shared embeddings and diffusion-specific embeddings, respectively. 
The deterministic-diffusion shared layer consists of several ResNet blocks and a T-RF bin processing block. 
It performs downsampling only along the frequency axis to preserve the deterministic information.
The deterministic-specific layer consists of several ResNet blocks. 
The attention layer is also applied to some layers, but not all. 
It performs downsampling either along the time axis or the time-frequency axis. 

The bottleneck processing module extracts the embedding for score estimation: 
\begin{equation}
  \mathbf{e'_s} = \text{Emb}_\text{score}(\mathbf{e'_{ds}}, t)
\label{rdm_bottle}
\end{equation}
$\text{Emb}_\text{score}$ represents the neural networks for processing embeddings in score estimation.

It should be noted that not all encoder layers were used for incorporating the deterministic information. 
As a result, the diffusion decoder and the deterministic decoder operate on different encoder layers. 
The deterministic-diffusion shared and deterministic-specific embeddings, along with skip connections, are processed separately by each decoder to obtain the final score and deterministic features: 
\begin{equation}
\left\{
\begin{aligned}
   \mathbf{est_{deter}} & = \text{Dec}_\text{deter}(\mathbf{e'_{dd}}, t) \\
   \mathbf{est_{score}} & = \text{Dec}_\text{score}(\mathbf{e'_{s}}, \mathbf{e'_{dd}}, \mathbf{e'_{ds}}, t)
\end{aligned}
\right.
\label{rdm_decoder}
\end{equation}
where $\mathbf{est_{score}}$ and $\mathbf{est_{deter}}$ represent the score for diffusion and the estimated deterministic feature, respectively. 
The training loss of the network is the sum of the score error and the complex spectrogram-based MSE: 
\begin{equation}
    \mathcal{L}_{RDM} = \mathcal{L}_{Score} + \mathcal{L}_{Deter}
    \label{loss:rdm}
\end{equation}

To further enhance D2U-SE, we propose utilizing both deterministic-only and deterministic-noisy features as conditions. 
Fig.~\ref{fig:dual-streaming} shows the structure of the D3U-SE. 
Different from D2U-SE, both deterministic-noisy and deterministic-only features are used as conditions. 
\begin{equation}
\begin{cases}
   \mathbf{e'_{da}} = \text{Enc}^{a}_\text{DD}(\mathbf{X'}_t, \mathbf{X'}, t) \\ 
   \mathbf{e'_{db}} = \text{Enc}^{b}_\text{DD}(\mathbf{X'}_t, \mathbf{X'}, \mathbf{Y}, t)
\end{cases}
\end{equation}
where $\text{Enc}^{a}_\text{DD}$ and $\text{Enc}^{b}_\text{DD}$ represent the deterministic-diffusion shared layers with deterministic-only and deterministic-noisy conditions. 
Another T-RF bin Processing Block is used to fuse $\mathbf{e'_{da}}$ and $\mathbf{e'_{db}}$ as $\mathbf{e'_{dd}}$. 
The diffusion-specific layers, bottleneck processing module, and deterministic decoder follow the same processing as described in Equation~(\ref{rdm_encoder}), Equation~(\ref{rdm_bottle}), and Equation~(\ref{rdm_decoder}). 
The score estimation decoder is processed as follows: 
\begin{equation}
\begin{aligned}
   \mathbf{est_{score}} & = \text{Dec}_\text{score}(\mathbf{e'_{s}}, \mathbf{e'_{dd}}, \mathbf{e'_{ds}}, \mathbf{e'_{da}}, \mathbf{e'_{db}}, t)
\end{aligned}
\label{rdm_decoder}
\end{equation}
The loss function is the same as that in Equation~(\ref{loss:rdm}).

\begin{table*}[]
\renewcommand{\arraystretch}{1.2}
\caption{Performance of the proposed COFFEE in the Evaluation Sets of the CHiME4 dataset. \\}
\centering
\begin{tabular}{c|c|ccccc|c|cc|cccccc}
\toprule
\multirow{2}{*}{\textbf{ID.}} &
  \multirow{2}{*}{\textbf{T-E}} &
  \multicolumn{5}{c|}{\textbf{T-RF}} &
  \multirow{2}{*}{\textbf{T-F}} &
  \multirow{2}{*}{\textbf{\begin{tabular}[c]{@{}c@{}}\textbf{\#Params} \\ \textbf{(M)}\end{tabular}}} &
  \multirow{2}{*}{\textbf{\begin{tabular}[c]{@{}c@{}}\textbf{Macs} \\ \textbf{(G/s)}\end{tabular}}} &
  \multirow{2}{*}{\textbf{C\textsubscript{sig} $\uparrow$}} &
  \multirow{2}{*}{\textbf{C\textsubscript{bak} $\uparrow$}} &
  \multirow{2}{*}{\textbf{C\textsubscript{ovl} $\uparrow$}} &
  \multirow{2}{*}{\textbf{\begin{tabular}[c]{@{}c@{}}\textbf{SDR} \\ \textbf{(dB) $\uparrow$} \end{tabular}}} &
  \multirow{2}{*}{\textbf{W-P $\uparrow$}} &
  \multirow{2}{*}{\textbf{STOI $\uparrow$}} \\
  \cline{3-7}
  &
  &
  \textit{\textbf{D\textsubscript{1}}} &
  \textit{\textbf{D\textsubscript{2}}} &
  \textit{\textbf{D\textsubscript{3}}} &
  \textit{\textbf{D\textsubscript{4}}} &
  \textit{\textbf{D\textsubscript{5}}} &
  & & & & & & & & \\
  \midrule
  
  0 & \ding{51} & & & & & & & 3.8 & 5.9
     & 3.75 & 2.77 & 2.99 
     & 11.25 & 2.26 & 0.944
     \\
     \midrule

  1 & \ding{51} 
     & \ding{51} & & & & 
     & 
     & 4.4  & 6.4
     & 3.77 & 2.78 & 3.01 
     & 11.29 & 2.27 & 0.944
     \\

  2 & \ding{51} 
     & \ding{51} & \ding{51} & & & 
     & 
     & 5.1 & 7.1
     & 3.78 & 2.78 & 3.02 
     & 11.33 & 2.28 & 0.947
     \\

  3 & \ding{51} 
     & \ding{51} & \ding{51} & \ding{51} & & 
     & 
     & 5.6 & 8.4
     & 3.82 & 2.80 & 3.05 
     & 11.41 & 2.31 & 0.947
     \\

  4 & \ding{51} 
     & \ding{51} & \ding{51} & \ding{51} & \ding{51} & 
     & 
     & 6.1 & 10.5
     & 3.81 & 2.82 & 3.05 
     & 11.65 & 2.30 & 0.951
     \\

  5 & \ding{51} 
     & \ding{51} & \ding{51} & \ding{51} & \ding{51} & \ding{51}
     & 
     & 6.6 & 14.5
     & 3.88 & 2.88 & 3.12 
     & 12.05 & 2.38 & 0.954
     \\
     \midrule

   6 & \ding{51} 
     & & \ding{51} & & & 
     & 
     & 4.4 & 6.7
     & 3.77 & 2.78 & 3.00 
     & 11.35 & 2.26 & 0.947
     \\    

   7 & \ding{51} 
     & & & \ding{51} & & 
     & 
     & 4.3 & 7.2
     & 3.78 & 2.77 & 3.01 
     & 11.38 & 2.26 & 0.948
     \\

   8 & \ding{51} 
     & & & & \ding{51} & 
     & 
     & 4.3 & 8.1
     & 3.82 & 2.84 & 3.06 
     & 11.70 & 2.32 & 0.950
     \\

   9 & \ding{51} 
     & & & & & \ding{51}
     & 
     & 4.2 & 9.9
     & 3.85 & 2.84 & 3.08 
     & 11.83 & 2.34 & 0.952
     \\ 
     \midrule

     
   10 & \ding{51} 
     & & & & & \ding{51}
     & \ding{51}
     & 4.9 & 20.9
     & \textbf{\color{blue}3.90} & \textbf{\color{blue}2.89} & \textbf{\color{blue}3.14} & \textbf{\color{blue}12.21} & \textbf{\color{blue}2.40} & \textbf{\color{blue}0.956}
     \\

     \midrule

    \multicolumn{8}{c|}{Mixture} & - & -
    & 2.61 & 1.92 & 1.88 
    & 7.54 & 1.27 & 0.870
    \\
     
     \multicolumn{8}{c|}{DCCRN} & 3.7 & 5.6 & 3.58 & 2.66 & 2.82 & 11.12 & 2.10 & 0.939 \\

     \multicolumn{8}{c|}{DPRNN} & 2.7 & 85.5 & 3.73 & 2.76 & 2.98 & 11.30 & 2.27 & 0.942 \\

     \multicolumn{8}{c|}{TF-GridNet}  & 9.4 & 153.1 & 3.97 & 2.95 & 3.19 & 12.57 & 2.41 & 0.957 \\

\bottomrule
  
\end{tabular}
\label{table:trf_tf_blocks}
\end{table*}

\section{Experiments}
\label{experimental}
\subsection{Evaluation Datasets}
We used CHiME4 dataset\footnote{https://www.chimechallenge.org/challenges/chime4/index} 
to evaluate the proposed method. 
All the Channel~1~--~Channel~6 simulated data from the training set were used for training the models. 
For evaluation, only the Channel~5 data was used. 
All data had a 16 kHz sample rate. 


\subsection{Settings for the Baseline Models}
The DCCRN \cite{hu20g_interspeech}, Conv-TasNet \cite{8707065}, DualPathRNN (DPRNN) \cite{9054266}, and TF-GridNet \cite{10214650} were selected as the deterministic baselines. 
Except for TF-GridNet, all deterministic models were trained for 200 epochs. 
TF-GridNet was trained for only 50 epochs due to its high training cost. 
\begin{enumerate}[fullwidth]
    \item[-] \textbf{DCCRN}: 
    The model used a two-layer complex LSTM with 256 units per layer. 
    The STFT parameters included the window length of 512, the hop size of 128, and the FFT length of 512, with the Hann window function. 
    The convolution kernel size was 5. 
    The number of kernels in each encoder layer was [32, 64, 128, 256, 256, 256], while the number of kernels in each decoder layer was [512, 512, 512, 256, 128, 64]. 
    \item[-] \textbf{DPRNN}: 
    The number of expected features in the input was 256, and the number of features in the hidden state was 64. 
    The hidden size of the BLSTM was 128. 
    The kernel sizes for both the encoder and decoder were 2. 
    Layer normalization was applied, with 6 layers in total. 
    The chunk length was 250. 
    \item[-] \textbf{TF-GridNet}: 
    The embedding dimension was 128, and the number of hidden units in the LSTM was 200. 
    The self-attention mechanism had 4 heads. 
    The approximate dimensions of the frame-level key and value tensors were 512. 
    The kernel size for unfolding and DeConv1D was 3, with the hop size of 1. 
    The model included 4 GridNetBlocks. 
    The STFT parameters included the window length of 512, the hop size of 128, and the FFT length of 512, using the Hann window function.     
\end{enumerate}

All baseline diffusion models used the same noise scenarioal Score Network (NCSN++) architecture (the number of parameters was 65.6 million (M)) as \textbf{SGMSE+} \cite{10149431}. 
The real and imaginary parts of the complex spectrograms were used as input features. 
Only the deterministic enhanced features were used as the condition for diffusion, referred to as \textbf{Deter-Diffusion}. 
We compared \textbf{StoRM} (the 65.6M NCSN++ architecture was also used). 
Different from the original paper \cite{10180108}, the deterministic models for StoRM in this study used different structures instead of NCSN++. 
For \textbf{GP-Unified} \cite{10448429}, all encoder and decoder settings were the same as SGMSE+, resulting in 106M training parameters and 65.6M decoding parameters. 
All diffusion models were trained for 150 epochs.

\subsection{Settings for the Proposed Methods}
For the proposed deterministic method, the STFT parameters included the window length of 512, the hop size of 128, and the FFT length of 512, with the Hann window function. 
The settings for the encoder and decoder layers were the same as those in DCCRN. 
The embedding dimension within the T-E, T-RF, and T-F processing blocks was 48. 
The input and output dimensions of the intra- and inter-LSTM within the T-E and T-RF processing blocks were 144 and 96, respectively. 
The input and output dimensions of the intra- and inter-LSTM within the T-E processing blocks were 768 and 48, respectively. 
Additional specific settings are provided in the experimental section. 
For the proposed \textbf{D2U-SE} model, 3 deterministic-diffusion shared encoder layers with 1 T-RF processing block were used; 5 diffusion-specific encoder layers were used; 3 deterministic decoder layers and 8 diffusion decoder layers were used. 
For the proposed \textbf{D3U-SE} model, we reduced the number of channels in the deterministic-diffusion shared encoder layers by half to ensure that the performance improvement was not due to an increase in the number of parameters. 
We have released the source code for the proposed model at:  \url{https://github.com/hshi-speech/Repair-Diffusion}.

\begin{table*}[h!]
\renewcommand{\arraystretch}{1.2}
\caption{Performance of Deter-Diffusion, StoRM, GP-Unified, and the proposed RDM-SE with DCCRN, DPRNN, and TF-GridNet enhanced conditions on the evaluation sets of the CHiME4 dataset. 
$\text{D}_{c}$ Decoder represents the use of the deterministic decoder. 
\#Parameters represents the number of parameters. 
}
\centering
\begin{tabular}{l|cc|ccccccc|ccccc}
\toprule
\multirow{2}{*}{\textbf{Systems}} & \multicolumn{2}{c|}{\textbf{\#Parameters}} 
&
  \multicolumn{7}{c|}{\textbf{Simulated}} 
  &
  \multicolumn{5}{c}{\textbf{Real}} \\
  \cline{4-15}
   & \textbf{Train} 
   & \textbf{Inference} & 
  \textbf{C\textsubscript{sig}} &
  \textbf{C\textsubscript{bak}} &
  \textbf{C\textsubscript{ovl}} 
  &
  \textbf{SDR}
  &
  \textbf{W-P}
  &
  \textbf{STOI} &
  \textbf{dMOS} 
  & \textbf{C\textsubscript{sig}} 
  & \textbf{C\textsubscript{bak}} 
  & \textbf{C\textsubscript{ovl}} 
  & \textbf{dMOS} & 
  \textbf{uMOS} 
  \\

  \midrule
   Mixture & - & -
    & 2.61 & 1.92 & 1.88 
    & 7.54 & 1.27 & 0.870
    & 2.69 
    & 2.08 & 1.47 & 1.46 & 2.47 & 1.58
    \\
  \midrule
   SGMSE+ & 65.6 & 65.6
    & 3.53 &  2.66 & 2.84 & 11.23 & 2.19 & 0.949 &  3.54 
    & 3.42 & 3.21 & 2.75
    & 3.45 & 2.63
    \\

  \midrule
   DCCRN & 3.7 & 3.7
   & 3.58 & 2.66 & 2.82 & 11.12 & 2.10 & 0.939 & 3.26 & 3.16 & 3.82 & 2.80 & 3.25 & 2.59 \\
   \quad + Deter-Diffusion & 69.3 & 69.3
   & 3.45 & 2.73 & 2.77 & 11.70 & 2.13 & 0.945 & 3.63 & 3.25 & 4.00 & 2.97 & 3.66 & 2.93 \\
   \quad + StoRM & 69.3 & 69.3 & 3.56 & 2.73 & 2.87 & 11.81 & 2.22 & 0.948 & 3.60 & 3.45 & 3.81 & 3.05& 3.57 & 3.16 \\
   \quad + GP-Unified & 109.1 & 69.3 & 3.50 & 2.71 & 2.82 & 11.92 & 2.17 & 0.949 & 3.58 & 3.46 & 3.81 & 3.06 & 3.58 & 3.28 \\
   \quad + D2U-SE & 25.2 & 25.2 & 3.58 & 2.75 & 2.88 & 11.65 & 2.22 & 0.950 & 3.61 & 3.43 & 4.00 & 3.12 & 3.67 & 3.46 \\
   \quad \quad + $\text{D}_{c}$ Decoder & 31.2 & 25.2 & 3.53 & 2.74 & 2.85 & 11.60 & 2.21 & 0.950 & 3.64 & 3.45 & 4.02 & 3.15 & 3.70 & 3.45 \\
   \quad + D3U-SE & 26.1 & 24.0 & 3.48 & 2.74 & 2.81 & 11.50 & 2.17 & 0.949 & \textbf{\color{blue}3.66} & 3.40 & 4.08 & 3.13 & \textbf{\color{blue}3.73} & \textbf{\color{blue}3.49} \\ 
   
  \midrule

   DPRNN & 2.7 & 2.7 
   & 3.73 & 2.76 & 2.98 & 11.30 & 2.27 & 0.942 & 3.34 & 3.39 & 3.94 & 3.05 & 3.07 & 2.90 \\
   \quad + Deter-Diffusion & 68.3 & 68.3 
   & 3.52 & 2.74 & 2.84 & 11.31 & 2.20 & 0.939 & 3.65 & 3.35 & 4.03 & 3.06 & 3.65 & 3.36 \\
   \quad + StoRM & 68.3 & 68.3 
   & 3.58 & 2.74 & 2.90 & 11.66 & 2.26 & 0.945 & 3.61 & 3.48 & 3.78 & 3.07 & 3.56 & 3.32 \\
   \quad + GP-Unified & 108.1 & 68.3 & 3.62 & 2.78 & 2.94 & 11.72 & 2.31 & 0.948 & 3.66 & 3.49 & 3.80 & 3.09 & 3.61 & 3.42 \\
   \quad + D2U-SE & 24.2 & 24.2 & 3.64 & 2.82 & 2.96 & 11.97 & 2.32 & 0.948 & 3.64 & 3.43 & 3.92 & 3.09 & 3.62 & 3.43 \\
   \quad \quad + $\text{D}_{c}$ Decoder & 30.2 & 24.2 & 3.58 & 2.80 & 2.92 & 11.87 & 2.30 & 0.948 & 3.63 & 3.43 & 3.91 & 3.09 & 3.61 & 3.45 \\
   \quad + D3U-SE & 25.1 & 23.0 & 3.56 & 2.80 & 2.90 & 11.82 & 2.28 & 0.948 & \textbf{\color{blue}3.66} & 3.40 & 4.03 & 3.12 & \textbf{\color{blue}3.69} & \textbf{\color{blue}3.53} \\

\midrule

   TF-GridNet & 9.4 & 9.4 
   & 3.97 & 2.95 & 3.19 & 12.57 & 2.41 & 0.957 & 3.55 & 3.11 & 4.00 & 2.81 & 3.35 & 2.71 \\
   \quad + Deter-Diffusion & 75.0 & 75.0 
   & 3.46 & 2.72 & 2.79 & 11.17 & 2.18 & 0.939 & 3.60 & 3.37 & 4.02 & 3.08 & 3.56 & \textbf{\color{blue}3.34} \\
   \quad + StoRM & 75.0 & 75.0 
   & 3.69 & 2.86 & 3.01 & 12.36 & 2.35 & 0.954 & 3.67 & 3.27 & 3.89 & 2.93 & 3.52 & 2.89 \\
   \quad + GP-Unified & 114.8 & 75.0 & 3.68 & 2.84 & 2.99 & 12.39 & 2.34 & 0.954 & 3.65 & 3.29 & 3.89 & 2.94 & 3.52 & 2.96 \\
   \quad + D2U-SE & 30.9 & 30.9 & 3.60 & 2.82 & 2.93 & 12.30 & 2.28 & 0.951 & 3.63 & 3.19 & 4.01 & 2.90 & 3.47 & 2.95 \\
   \quad \quad + $\text{D}_{c}$ Decoder & 36.9 & 30.9 & 3.68 & 2.85 & 2.98 & 12.28 & 2.32 & 0.952 & 3.63 & 3.17 & 4.01 & 2.89 & 3.51 & 3.02 \\
   \quad + D3U-SE & 31.8 & 29.7 & 3.50 & 2.82 & 2.84 & 12.40 & 2.21 & 0.952 & \textbf{\color{blue}3.70} & 3.21 & 4.09 & 2.95 & \textbf{\color{blue}3.58} & 3.09 \\

   \midrule
   COFFEE & 4.9 & 4.9 & 3.90 & 2.89 & 3.14 & 12.21 & 2.40 & 0.956 & 3.43 & 3.40 & 3.98 & 3.08 & 3.47 & 3.20 \\
   \quad + Deter-Diffusion & 70.5 & 70.5 & 3.63 & 2.85 & 2.95 & 11.96 & 2.31 & 0.950 & 3.68 
    & 3.30 & {4.07} & 3.04
    & 3.75 & 3.23 \\
   \quad + StoRM & 70.5 & 70.5 & {3.73} & {2.88} & {3.04} & 12.20 & {2.38} & {0.954} & \textbf{\color{blue}3.72} 
    & 3.45 & 4.02 & 3.15
    & \textbf{\color{blue}3.80} & 3.43 \\
   \quad + GP-Unified & 110.3 & 70.5 & 3.69 & 2.85 & 3.00 & {12.21} & 2.35 & 0.954 & 3.70 
    & 3.48 & 3.97 & 3.15
    & 3.72 & 3.47 \\
   \quad + D2U-SE & 26.4 & 26.4 & 3.67 & 2.85 & 2.99 & 11.95 & 2.34 & 0.953 & 3.67 & 3.42 & 4.07 & 3.15 & 3.75 & 3.58 \\
   \quad \quad + $\text{D}_{c}$ Decoder & 32.4 & 26.4 & {3.69} & {2.86} & {3.01} & {11.87} & {2.37} & {0.954} & {3.68} & {3.44} & {4.06} & {{3.17}} & {3.76} & {3.58} \\
   \quad + D3U-SE & 27.3 & 25.2 
   & 3.65 & 2.85 & 2.98 & 12.01 & 2.34 & 0.954
   & 3.69 & 3.44 & 4.09 & 3.18 & 3.77 & \textbf{\color{blue}3.60}\\

\bottomrule
\end{tabular}
\label{table:sgmse_storm}
\end{table*}

\subsection{Additional Settings for Evaluation}
\label{sec:evalution_desc}
In the diffusion process, we used PC sampling with 30 steps. 
In previous work \cite{shiensemble}, we found that within a 30-step reverse diffusion process, the sampling diversity at step 20 (with $t$ progressing from $T$ to $0$) can be maintained as if starting from the initial state. 
Thus, in this paper, we start the diffusion process at $t=20$ instead of $t=30$. 
In previous work \cite{shiensemble}, we found that when the ensemble number was greater than 8, the performance difference became negligible. 
Therefore, this paper used the ensemble with a maximum of 8 samples. 

We test for statistical significance using a two-sided paired $t$-test with $n=330$ (real) and $n=420$ (simulated) for uMOS/dMOS. 
With sample standard deviation of paired differences $s_{\Delta}=0.50$, the standard error is $\mathrm{SE}=s_{\Delta}/\sqrt{n}\approx 0.0275$ (real) and $0.0244$ (simulated). 
Hence an average gain $\Delta\mathrm{MOS}\ge 0.10$ is statistically significant (its 95\% confidence interval excludes $0$) for both sets.

\subsection{Evaluation Metrics}
Multiple evaluation metrics were used to analyze the improvements in human listening experience. 
For the simulated data, signal distortion ($\text{C\textsubscript{sig}}$) \cite{4389058}, background intrusiveness ($\text{C\textsubscript{bak}}$) \cite{4389058}, overall quality ($\text{C\textsubscript{ovl}}$) \cite{4389058}, the wideband perceptual evaluation of speech quality (W-P) \cite{941023}, the short-time objective intelligibility (STOI) \cite{5495701}, and signal to distortion ratio (SDR) were used as evaluation metrics. 
For the real data, the Mean Opinion Score was used, which was evaluated with the open-source toolkit DNSMOS (dMOS) \cite{reddy2021dnsmos,reddy2022dnsmos}. 
UTMOS (uMOS) \cite{saeki22c_interspeech} was also used. 
Although both the dMOS and uMOS are designed to approximate subjective perception, they still remain objective metrics.

\subsection{Evaluation in the Proposed Deterministic Model}
Table~\ref{table:trf_tf_blocks} shows the performance of the proposed COFFEE method. 
We compared the effects of T-RF and T-F across different layers. 
Exp.--1 to Exp.--10 compared the impact of using T-RF in different decoder layers. 
From the first to the fifth decoder layer, simply stacking the network in different layers improved all evaluation metrics. 
Through Exp.--6 to Exp.--9, we compared the impact of T-RF in different decoder layers. 
The experimental results show that applying fine-grained processing closer to the output layer led to more significant performance improvements. 
Comparing Exp.--5 and Exp.--9, although Exp.--5 showed better performance, using fine-grained processing only in the final decoder layer provided advantages in terms of the number of parameters and processing speed, with minimal performance difference between these two models. 
Exp.--10 used a T-F processing block after the decoder. 
Compared to using T-RF blocks in every decoder layer (Exp.--5), using fine-grained processing only after the final decoder layer or after the decoder yielded better performance. 
We compared the proposed method with other deterministic models. 
TF-GridNet is currently the best-performing model on the CHiME4 dataset. 
The proposed model had comparable performance to TF-GridNet. 
The proposed COFFEE outperformed DCCRN and DPRNN.

\subsection{Evaluation of Different Diffusion Models}
Table~\ref{table:sgmse_storm} shows the performance of different deterministic models in combination with various diffusion models. 
Compared to the noisy mixture signal, SGMSE+ effectively enhances noisy signals. 
Deter-Diffusion, which used only deterministic enhanced features as conditions, performed better than SGMSE+. 
When using DCCRN-enhanced features as conditions, the performance of objective evaluation metrics ($\text{C\textsubscript{sig}}$, $\text{C\textsubscript{bak}}$, $\text{C\textsubscript{ovl}}$, SDR, WB-PESQ, STOI) on the simulated dataset was comparable to that of SGMSE+. 
This may be due to the signal distortions introduced by DCCRN. 
The DCCRN-Deter-Diffusion performed better at noise reduction, achieving the higher $\text{C\textsubscript{bak}}$ value, while the noisy SGMSE+ method excelled at speech signal recovery, yielding the higher $\text{C\textsubscript{sig}}$ value. 
Although DCCRN-Deter-Diffusion did not outperform SGMSE+ in simulated objective evaluation metrics, it showed certain advantages in MOS: Deter-Diffusion achieved higher scores on both dMOS and uMOS on the real evaluation sets. 
The quality of the deterministic enhanced features affected the final performance of Deter-Diffusion. 
The performance ranking of the deterministic models on the simulated evaluation sets: TF-GridNet $>$ DPRNN $>$ DCCRN. 
However, TF-GridNet-Deter-Diffusion performed the worst on the simulated data. 
While the performance on the real evaluation sets was inconsistent with that on the simulated evaluation sets: DPRNN $>$ TF-GridNet $>$ DCCRN. 
TF-GridNet outperformed DPRNN in dMOS but underperformed in uMOS. 
The TF-GridNet-Deter-Diffusion and DPRNN-Deter-Diffusion performed worse in dMOS on the real evaluation sets than the DCCRN-Deter-Diffusion. 

The Deter-Diffusion models did not outperform the corresponding deterministic models in certain objective evaluation metrics ($\text{C\textsubscript{sig}}$, $\text{C\textsubscript{bak}}$, $\text{C\textsubscript{ovl}}$, WB-PESQ, STOI) on the simulated evaluation sets, but they exhibited better MOS values on the real evaluation sets. 
This may be because the diffusion model tends to capture the data distribution rather than frame-level signal restoration. 
As a result, improving objective evaluation metrics becomes challenging, but the model is able to obtain features that are closer to the clean speech data distribution.

We also compared two models that incorporate the original noisy features as conditions: StoRM and GP-Unified. 
The models with deterministic-noisy conditions demonstrated greater sensitivity to the deterministic enhanced features, regardless of whether deterministic information was incorporated into the diffusion model implicitly or explicitly during training.

Compared to Deter-Diffusion, StoRM demonstrated advantages in objective evaluation metrics ($\text{C\textsubscript{sig}}$, $\text{C\textsubscript{bak}}$, $\text{C\textsubscript{ovl}}$, WB-PESQ, STOI) in most cases on the simulated evaluation sets. 
However, on the simulated evaluation sets, only the TF-GridNet enhanced features achieved better dMOS values, while using DCCRN and DPRNN enhanced features as conditions resulted in performance degradation. 
These conclusions differ from those observed on the real evaluation sets. 
With DCCRN and DPRNN enhanced conditions, StoRM achieved better signal recovery ($\text{C\textsubscript{sig}}$) and signal overall recovery ($\text{C\textsubscript{ovl}}$) compared to Deter-Diffusion. 
However, StoRM with TF-GridNet-enhanced condition had worse enhancement performance compared to Deter-Diffusion. 
Compared to Deter-Diffusion, all baseline deterministic enhanced conditions resulted in a decline in the noise suppression ($\text{C\textsubscript{bak}}$). 
All StoRM models resulted in lower dMOS values. 
For uMOS, only the DCCRN-enhanced StoRM improved performance, while DPRNN-enhanced and TF-GridNet-enhanced conditions led to a decline compared to Deter-Diffusion.

GP-Unified explicitly incorporates deterministic information by using a deterministic decoder during training. 
Compared to StoRM, GP-Unified had comparable performance on the simulated evaluation sets but had better performance on the real evaluation sets. 
For all dMOS and uMOS values on the real evaluation sets, GP-Unified outperformed StoRM. 
However, compared to Deter-Diffusion, GP-Unified still showed a decline in some metrics. 
Only with TF-GridNet-enhanced conditions, GP-Unified achieved better performance on the simulated evaluation sets (WB-PESQ, STOI). 
For the real evaluation sets, GP-Unified still had some disadvantages in dMOS values, which were similar to those of StoRM: with DCCRN or DPRNN conditions, both StoRM and GP-Unified had a decline in dMOS values. 
By comparing Deter-Diffusion, StoRM, and GP-Unified across various metrics on the real evaluation sets, we found significant declines in noise reduction ($\text{C\textsubscript{bak}}$) for StoRM and GP-Unified. 
This may explain the decrease in dMOS values: \cite{kaneko24_interspeech} indicates that the uMOS is insensitive to speech with noise compared to dMOS. 
For uMOS values, with DCCRN or DPRNN conditions, GP-Unified outperformed Deter-Diffusion. 
However, with TF-GridNet conditions, although it showed some improvement over StoRM, it still had significant performance degradation compared to Deter-Diffusion. 
Based on the performance of TF-GridNet as the condition on simulated and real evaluation sets, we argue that the T-F bin-level fine-grained processing, while achieving best performance on simulated data, may introduce distortions more easily to real data due to distribution differences between real and simulated data. 
However, the advantages brought by fine-level processing cannot be ignored for deterministic models. 
Besides, although DCCRN has lower dMOS and uMOS values than TF-GridNet in real evaluation sets, its performance was acceptable and demonstrates more stable diffusion performance, particularly in models that combine deterministic-noisy features.

\begin{figure*}[htbp]
  \centering
  \captionsetup[subfloat]{labelformat=empty}
  \makebox[\linewidth]{%
      \includegraphics[width=1.32\textwidth]{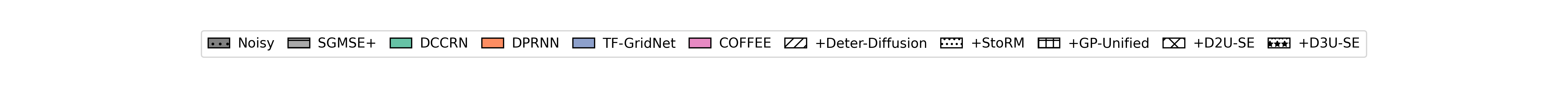}
  }
  \vspace{-40pt}
  \captionsetup[subfloat]{labelformat=parens,labelsep=space}

  \par\vspace{1em}  
  \subfloat[dMOS score on BUS noise scenario]{%
    \includegraphics[width=0.49\textwidth]{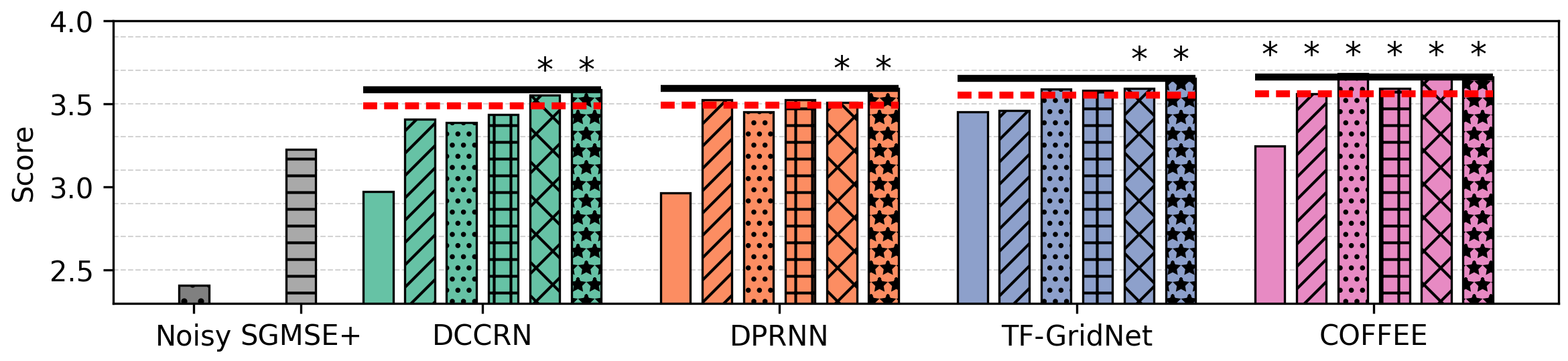}
    \label{fig:sub1}
  }
  \subfloat[dMOS score on  CAF noise scenario]{%
    \includegraphics[width=0.49\textwidth]{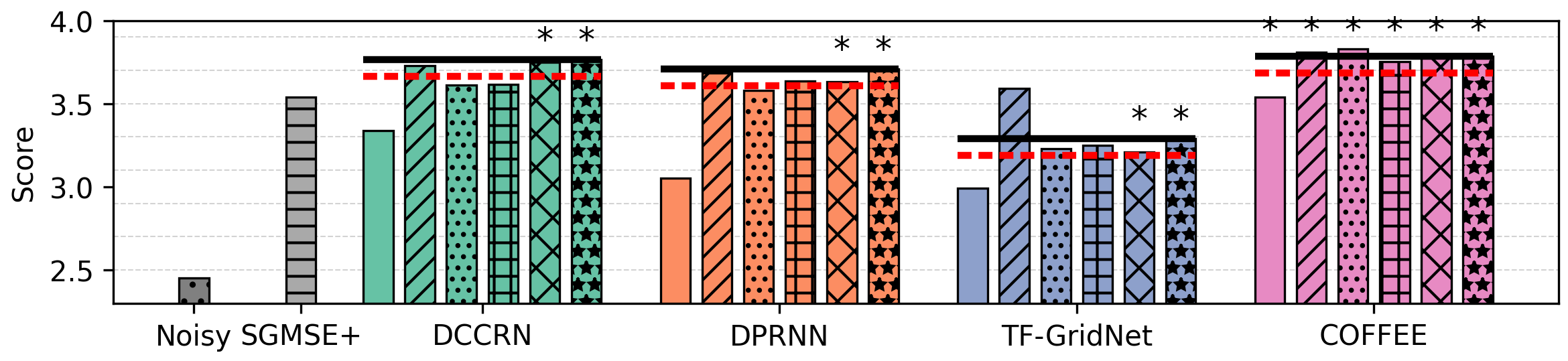}
    \label{fig:sub2}
  }
  \vspace{-20pt}
  \par\vspace{1em}  
  \subfloat[dMOS score on  PED noise scenario]{%
    \includegraphics[width=0.49\textwidth]{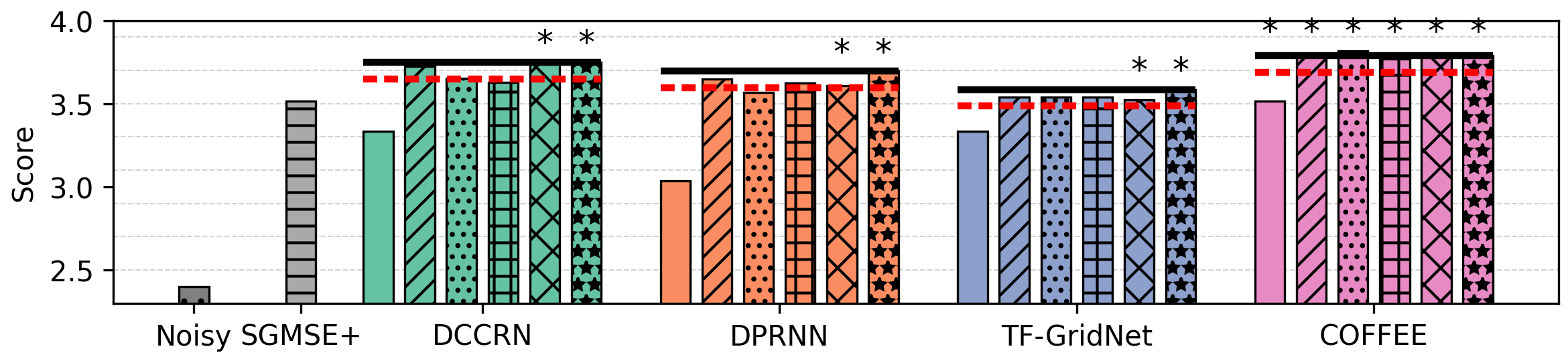}
    \label{fig:sub3}
  }
  \subfloat[dMOS score on  STR noise scenario]{%
    \includegraphics[width=0.49\textwidth]{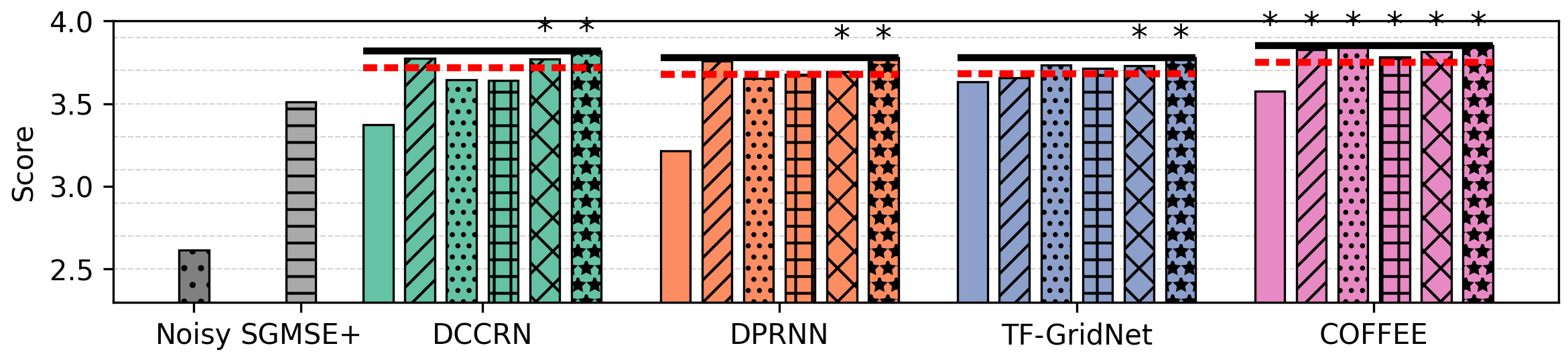}
    \label{fig:sub4}
  }

  \caption{dMOS scores under different noise scenarios using various SE systems on the real evaluation set of the CHiME4 dataset. 
  In each group, models below the red dashed line exhibit a statistically significant performance improvement (significance computed as described in Section~\ref{sec:evalution_desc}). 
  Bars marked with \protect\sixstar\ denote the proposed model.
  }
  \label{fig:dmos_conditions}
  \vspace{-10pt}
\end{figure*}

\begin{figure*}[htbp]
  \centering
  \captionsetup[subfloat]{labelformat=empty}
  \makebox[\linewidth]{%
      \includegraphics[width=1.32\textwidth]{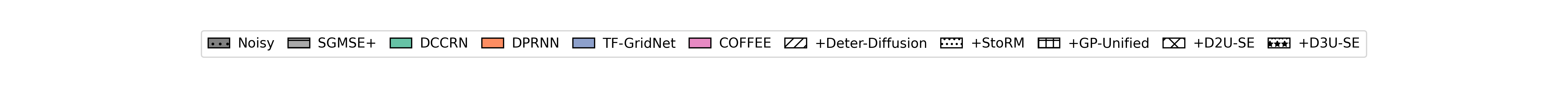}
  }
  \vspace{-40pt}
  \captionsetup[subfloat]{labelformat=parens,labelsep=space}

  \par\vspace{1em}  
  \subfloat[uMOS score on BUS noise scenario]{%
    \includegraphics[width=0.49\textwidth]{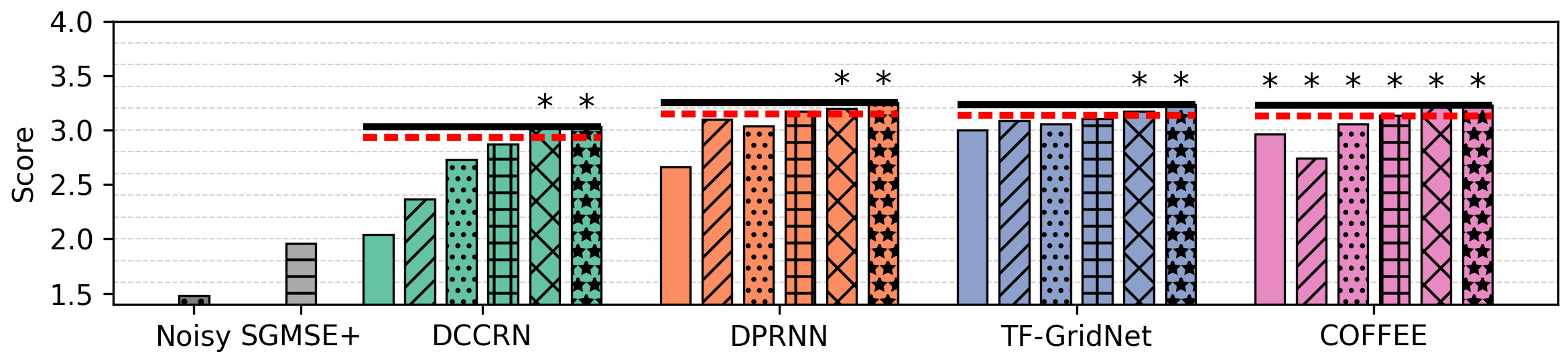}
    \label{fig:sub1}
  }
  \subfloat[uMOS score on CAF noise scenario]{%
    \includegraphics[width=0.49\textwidth]{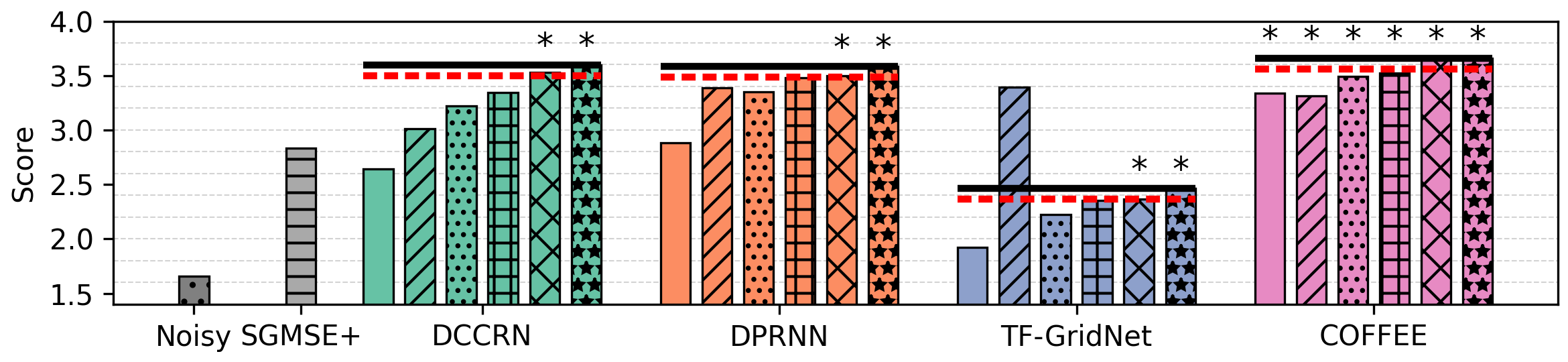}
    \label{fig:sub2}
  }
  \vspace{-20pt}
  \par\vspace{1em}  
  \subfloat[uMOS score on PED noise scenario]{%
    \includegraphics[width=0.49\textwidth]{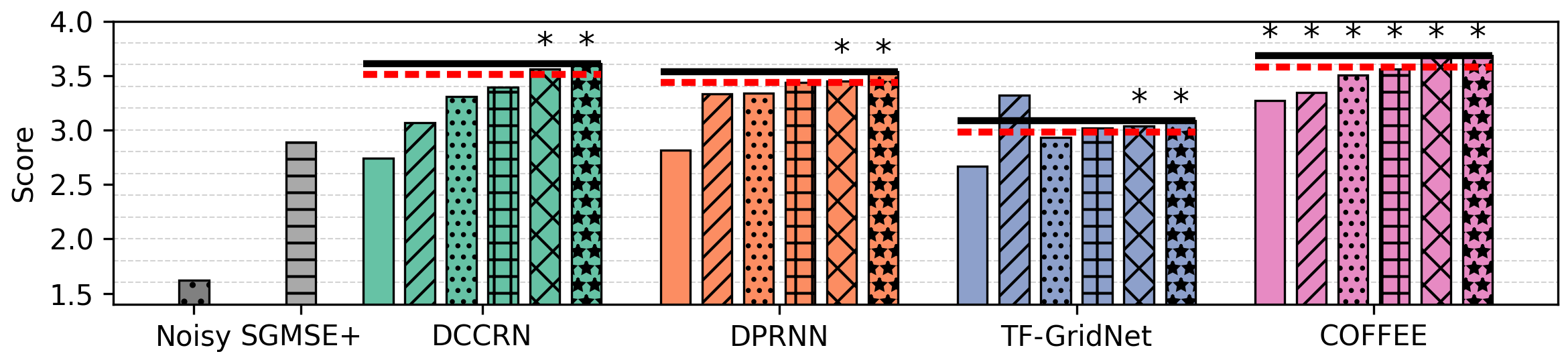}
    \label{fig:sub3}
  }
  \subfloat[uMOS score on STR noise scenario]{%
    \includegraphics[width=0.49\textwidth]{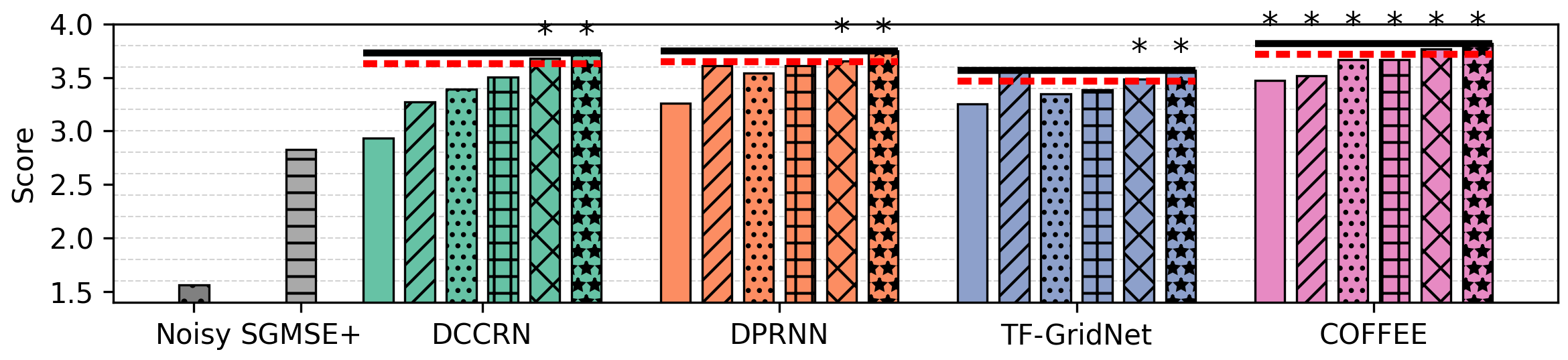}
    \label{fig:sub4}
  }

  \caption{uMOS scores under different noise scenarios using various SE systems on the real evaluation set of the CHiME4 dataset. 
  In each group, models below the red dashed line exhibit a statistically significant performance improvement (significance computed as described in Section~\ref{sec:evalution_desc}). 
  Bars marked with \protect\sixstar\ denote the proposed model.
  }
  \label{fig:umos_conditions}
\end{figure*}

Table~\ref{table:sgmse_storm} also shows our proposed D2U-SE. 
We also used DCCRN, DPRNN, and TF-GridNet as deterministic models for diffusion. 
Compared to StoRM and GP-Unified, the proposed D2U-SE achieved better and more stable performance. 
On the real evaluation sets, D2U-SE outperformed StoRM and GP-Unified in almost all deterministic models. 
Compared to StoRM and GP-Unified, D2U-SE achieved better noise suppression, particularly on the real evaluation sets, with higher $\text{C\textsubscript{bak}}$ values. 
D2U-SE showed a greater advantage in uMOS compared to StoRM and GP-Unified: using all deterministic models, D2U-SE achieved consistent improvements compared to other deterministic-noisy diffusion models. 
For dMOS, different deterministic models had varying impacts on D2U-SE: with DCCRN and DPRNN, D2U-SE outperformed StoRM and GP-Unified; while with TF-GridNet, D2U-SE had comparable performance to StoRM and GP-Unified. 
Compared to Deter-Diffusion, D2U-SE showed significant improvements when using DCCRN, comparable performance when using DPRNN, and a slight decline when using TF-GridNet. 
Surprisingly, D2U-SE with DCCRN outperformed those using DPRNN and TF-GridNet, achieving the best performance. 
This did not occur in StoRM or GP-Unified. 
As previously analyzed, fine-grained processing may result in more severe distortion issues when handling real data. 
This suggests that while the proposed D2U-SE can mitigate distortion issues to some extent, its performance is still closely tied to the input deterministic feature. 
Explicitly incorporating deterministic repair ($ + \text{D}_{c}$ Decoder) further improved the model's performance, especially when TF-GridNet was used as the condition. 
In addition, given the performance differences among different deterministic methods under deterministic-only conditions (Deter-Diffusion) and deterministic-noisy conditions (StoRM and GP-Unified), we proposed to use dual-stream encoding (D3U-SE). 
Although the dual-stream encoding model showed some performance degradation on simulated metrics ($\text{C\textsubscript{sig}}$, $\text{C\textsubscript{bak}}$, $\text{C\textsubscript{ovl}}$, SDR, WB-PESQ, STOI) compared to D2U-SE ($ + \text{D}_{c}$ Decoder), it achieved improvements in dMOS and uMOS.

\begin{table}[h!]
\renewcommand{\arraystretch}{1.}
\caption{The summary performance with Mean ($\pm$ Standard Deviation) of different methods under evaluation sets. 
}
\centering
\begin{tabular}{l|c|cc}
\toprule
\multirow{2}{*}{\textbf{Systems}}
&
  \multicolumn{1}{c|}{\textbf{Simulated}} 
  &
  \multicolumn{2}{c}{\textbf{Real}} \\
  \cline{2-4}
  & \textbf{dMOS} & \textbf{dMOS} & \textbf{uMOS}  \\

  \midrule
   Mixture 
    & 2.69 ($\pm$0.18) & 2.47 ($\pm$0.22) & 1.58 ($\pm$0.51)
    \\
  \midrule
   SGMSE+ 
    & 3.54 ($\pm$0.23)  & 3.45 ($\pm$0.31) & 2.63 ($\pm$0.65)
    \\

  \midrule
   DCCRN 
   & 3.26 ($\pm$0.22) & 3.25 ($\pm$0.30) & 2.59 ($\pm$0.61) \\
   \quad + Deter-Diffusion 
   & 3.63 ($\pm$0.20) & 3.66 ($\pm$0.30) & 2.93 ($\pm$0.59) \\
   \quad + StoRM 
   & 3.60 ($\pm$0.21) & 3.57 ($\pm$0.29) & 3.16 ($\pm$0.50) \\
   \quad + GP-Unified 
   & 3.58 ($\pm$0.20) & 3.58 ($\pm$0.26) & 3.28 ($\pm$0.45) \\
   \quad + D3U-SE 
   & 3.66 ($\pm$0.21) & 3.73 ($\pm$0.26) & 3.49 ($\pm$0.49) \\ 
   
  \midrule

   DPRNN 
   & 3.34 ($\pm$0.20) & 3.07 ($\pm$0.23) & 2.90 ($\pm$0.54) \\
   \quad + Deter-Diffusion 
   & 3.65 ($\pm$0.21) & 3.65 ($\pm$0.25) & 3.36 ($\pm$0.48) \\
   \quad + StoRM 
   & 3.61 ($\pm$0.21) & 3.56 ($\pm$0.25) & 3.32 ($\pm$0.48) \\
   \quad + GP-Unified 
   & 3.66 ($\pm$0.21) & 3.61 ($\pm$0.23) & 3.42 ($\pm$0.45) \\
   \quad + D3U-SE 
   & 3.66 ($\pm$0.21) & 3.69 ($\pm$0.25) & 3.53 ($\pm$0.48) \\

\midrule

   TF-GridNet 
   & 3.55 ($\pm$0.23) & 3.35 ($\pm$0.43) & 2.71 ($\pm$0.85) \\
   \quad + Deter-Diffusion 
   & 3.60 ($\pm$0.20) & 3.56 ($\pm$0.23) & 3.34 ($\pm$0.48) \\
   \quad + StoRM 
   & 3.67 ($\pm$0.20) & 3.52 ($\pm$0.39) & 2.89 ($\pm$0.75) \\
   \quad + GP-Unified 
   & 3.65 ($\pm$0.21) & 3.52 ($\pm$0.39) & 2.96 ($\pm$0.71) \\
   \quad + D3U-SE 
   & 3.70 ($\pm$0.21) & 3.58 ($\pm$0.39) & 3.09 ($\pm$0.75) \\

   \midrule
   COFFEE 
   & 3.43 ($\pm$0.21) & 3.47 ($\pm$0.29) & 3.20 ($\pm$0.48) \\
   \quad + Deter-Diffusion 
   & 3.68 ($\pm$0.19) & 3.75 ($\pm$0.26) & 3.23 ($\pm$0.55) \\
   \quad + StoRM 
   & 3.72 ($\pm$0.20) & 3.80 ($\pm$0.24) & 3.43 ($\pm$0.47) \\
   \quad + GP-Unified 
   & 3.70 ($\pm$0.20) & 3.72 ($\pm$0.26) & 3.47 ($\pm$0.44) \\
   \quad + D3U-SE 
   & 3.69 ($\pm$0.20) & 3.77 ($\pm$0.25) & 3.60 ($\pm$0.46) \\

\bottomrule
\end{tabular}
\label{table:methods_standard-deviation}
\end{table}

\begin{figure}
    \centering
    \includegraphics[width=1.\linewidth]{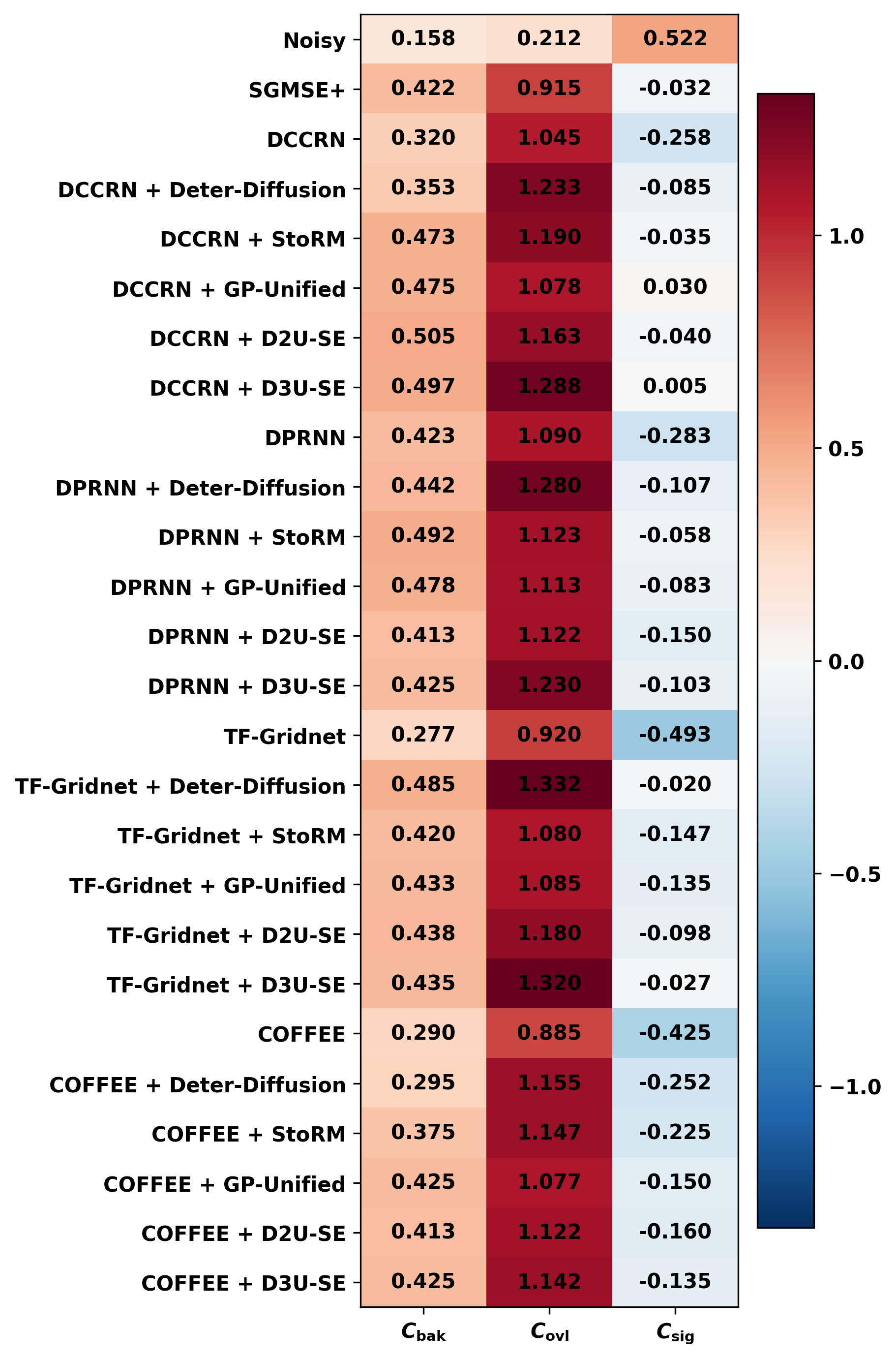}
    \caption{Heatmap showing the differences in $\text{C\textsubscript{sig}}$, $\text{C\textsubscript{bak}}$ and $\text{C\textsubscript{ovl}}$ scores when evaluated using DNSMOS and the Composite toolkit (requiring clean labels) under simulated evaluation sets. 
    The difference values are calculated as (DNSMOS estimated score $-$ Composite toolkit estimated score).}
    \label{fig:headmap}
\end{figure}

Table~\ref{table:sgmse_storm} shows the performance of the proposed deterministic model for diffusion on the evaluation sets of the CHiME4 dataset. 
Compared to other deterministic models, the proposed COFFEE achieved better dMOS and uMOS values on the real evaluation sets. 
On the real evaluation sets, Deter-Diffusion with the proposed COFFEE achieved better dMOS values compared to other baseline deterministic models. 
However, it only outperformed DCCRN for uMOS. 
Compared to other baseline deterministic models, Deter-Diffusion with COFFEE had better noise suppression ($\text{C\textsubscript{bak}}$). 
On the simulated evaluation sets, Deter-Diffusion with COFFEE achieved the best performance compared to other deterministic models. 
For StoRM and GP-Unified, using COFFEE as the deterministic model achieved better performance on both simulated and real evaluation sets compared to using the same diffusion model with other deterministic models. 
TF-GridNet with fine-grained processing did not achieve the best performance, while diffusion models with DCCRN demonstrated more stable performance, which motivated the design of COFFEE. 
Combining fine-grained and coarse-grained processing demonstrated strong performance: diffusion models with COFFEE outperformed those diffusion models using other deterministic models. 
In uMOS evaluations, diffusion models with COFFEE had the same trend as other deterministic models: D2U-SE $>$ GP-Unified $>$ StoRM. 
In addition, we also evaluated the standard deviations of different methods under both dMOS and uMOS, which is shown in Table~\ref{table:methods_standard-deviation}. 
The results show that, except when TF-GridNet was used as the condition—where the proposed D3U-SE experiences a performance drop compared to Deter-Diffusion—the performance improves while maintaining stability under other deterministic conditions, especially in terms of the uMOS metric.

Fig.~\ref{fig:dmos_conditions} and Fig.~\ref{fig:umos_conditions} compare the dMOS and uMOS performance of different SE models under various noise scenarios. 
Among all noise conditions, the BUS noise scenario was consistently the most challenging scenario for all models to recover effectively. 
In addition, dMOS appears to be more difficult to improve compared to uMOS: under the uMOS metric, the proposed D2U-SE and D3U-SE models showed statistically significant improvements in most cases when conditioned on different deterministic models. 
However, under the dMOS metric, significant improvements are observed only when using DCCRN or DPRNN as the conditioning models.
COFFEE demonstrates greater stability compared to other deterministic models: when used as the conditioning model, the dMOS scores across different diffusion models show no significant variation. 
When conditioned on TF-GridNet, the system showed greater instability in the CAF noise scenario: using the deterministic-only input significantly outperforms the deterministic-noisy condition. 
In contrast, this performance gap is much less pronounced in other noise environments. 

\subsection{Comparison of DNSMOS and Composite Toolkit Metrics}
In Table~\ref{table:sgmse_storm}, the $\text{C\textsubscript{sig}}$, $\text{C\textsubscript{bak}}$, and $\text{C\textsubscript{ovl}}$ values are computed using the Composite toolkit with the clean labels. 
The difference heatmap (\textit{\textit{DNSMOS estimated score  $-$ Composite toolkit estimated score}}) across the $\text{C\textsubscript{sig}}$, $\text{C\textsubscript{bak}}$, and $\text{C\textsubscript{ovl}}$ is shown in Fig.~\ref{fig:headmap}. 
For all enhancement systems, $\text{C\textsubscript{bak}}$ shows a positive gain, indicating that DNSMOS tends to underestimate the amount of noise in the enhanced signals. 
The overall quality score $\text{C\textsubscript{ovl}}$ shows a substantial overestimation of overall speech quality by DNSMOS. 
In contrast, the speech signal quality score $\text{C\textsubscript{sig}}$ is negative for most systems, indicating that DNSMOS underestimates speech quality. 
An interesting observation is that deterministic models generally yield smaller $\text{C\textsubscript{bak}}$ and $\text{C\textsubscript{ovl}}$ differences than diffusion models that use them as the corresponding condition, but larger differences in $\text{C\textsubscript{sig}}$. 
This observation suggests that, to achieve more stable performance—particularly in certain downstream tasks—it may be necessary to introduce a bias or set higher or lower thresholds for evaluation metrics like DNSMOS and UTMOS, relative to conventional objective measures.

\section{Conclusions}
\label{conclusions}
Conventional diffusion models need to incorporate accurate prior knowledge as reliable conditions to generate accurate predictions. 
In this paper, we propose a Deterministic-Diffusion Unified model for Speech Enhancement (D2U-SE) to provide reliable conditions for diffusion using the deterministic model. 
It explicitly incorporates deterministic information using an additional deterministic decoder. 
We first experimented with using DCCRN, DPRNN, and TF-GridNet as deterministic models to evaluate their impact on diffusion models on CHiME4 datasets. 
We found that while using diffusion models did not necessarily enhance objective metrics on simulated evaluation sets, it had certain advantages on real evaluation sets. 
The proposed D2U-SE was compared with Deter-Diffusion, which directly uses enhanced deterministic features as conditions, as well as with StoRM and GP-Unified, which use both noisy and enhanced deterministic features as conditions. 
The proposed D2U-SE had better noise suppression and further improved both dMOS and uMOS values compared to StoRM and GP-Unified. 
Deterministic models had a significant impact on diffusion performance. 
The experimental results also suggested that the choice between deterministic-only and deterministic-noisy conditions depends on the specific deterministic model. 
Thus, we further improved the D2U-SE with dual-stream encoding to utilize both deterministic-only and deterministic-noisy conditions (D3U-SE). 
Furthermore, deterministic models with fine-grained processing had unstable performance on real evaluation sets, whereas coarse-grained processing based on the UNet structure had more stable diffusion performance. 
Therefore, we proposed a deterministic model that combines coarse-grained and fine-grained processing, called COarse First then
Fine EnhancEment (COFFEE). 
With COFFEE as the deterministic model, different diffusion models had stable and superior performance compared to those using other deterministic models.

\bibliographystyle{IEEEtran}
\bibliography{mybib}

\end{document}